\documentclass[fleqn,usenatbib]{mnras}
\usepackage{mathptmx}
\usepackage[T1]{fontenc}
\DeclareRobustCommand{\VAN}[3]{#2}
\let\VANthebibliography\thebibliography
\def\thebibliography{\DeclareRobustCommand{\VAN}[3]{##3}\VANthebibliography}
\usepackage{graphicx}	
\usepackage{amsmath}	
\usepackage{amssymb}	
\usepackage{rotating}
\usepackage{sidecap}
\usepackage{lscape}
\usepackage{multicol, blindtext}

\usepackage{aecompl}
\usepackage{wasysym}
\usepackage{subcaption}
\usepackage{array}
\usepackage{times}
\usepackage{url}
\usepackage{xcolor}
\usepackage{hyperref}
\hypersetup{colorlinks=true,citecolor=blue,linkcolor=blue,filecolor=blue,urlcolor=blue}
\usepackage{soul}

\title[Cluster galaxy evolution with residence time]{How time weathers galaxies: The temporal impact of the cluster environment on galaxy formation and evolution}
\author[S.~O'Neil et al.]
    {{Stephanie O'Neil$^{1}$ \thanks{E-mail: sloneil@mit.edu},
    Josh Borrow$^{1}$,
    Mark Vogelsberger$^{1,2}$,
    Hanzhang Zhao$^{3}$,
    Bing Wang$^{4}$}\\
    $^{1}$Department of Physics and Kavli Institute for Astrophysics and Space Research,
           Massachusetts Institute of Technology,
           Cambridge, MA 02139, USA\\
    $^{2}$The NSF AI Institute for Artificial Intelligence and Fundamental Interactions, Massachusetts Institute of Technology, Cambridge, MA 02139, USA\\
    $^{3}$Department of Physics and Astronomy, Texas A\&M University, College Station, Texas 77843, USA\\
    $^{4}$ Department of Medical Oncology, Amsterdam University Medical Center, Cancer Center Amsterdam, Vrije Universiteit, 1081 HV Amsterdam,\\ the Netherlands\\
    }
\begin{document}

\date{Accepted 2024 April 04. Received 2024 April 04; in original form 2023 September 28}

\pagerange{\pageref{firstpage}--\pageref{lastpage}}
\pubyear{2023}

\maketitle

\label{firstpage}

\begin{abstract}
We illuminate the altered evolution of galaxies in clusters compared to central galaxies by tracking galaxies in the IllustrisTNG300 simulation as they enter isolated clusters of mass $10^{13} < M_{\rm 200, mean} / {\rm M}_\odot < 10^{15}$ (at $z=0$).
We demonstrate significant trends in galaxy properties with residence time (time since first infall) and that there is a population of galaxies that remain star-forming even many Gyrs after their infall.
By comparing the properties of galaxies at their infall time to their properties at $z=0$, we show how scaling relations, like the stellar-to-halo mass ratio, shift as galaxies live in the cluster environment. 
Galaxies with a residence time of 10 Gyr increase their stellar-to-halo mass ratio, by around 1 dex.
As measurements of the steepest slope of the galaxy cluster number density profile ($R_{\rm st}$), frequently used as a proxy for the splashback radius, have been shown to depend strongly on galaxy selection, we show how $R_{\rm st}$ depends on galaxy residence time.
Using galaxies with residence times less than one cluster crossing time ($\approx 5$ Gyr) to measure $R_{\rm st}$ leads to significant offsets relative to using the entire galaxy population. Galaxies must have had the opportunity to `splash back' to the first caustic to trace out a representative value of $R_{\rm st}$, potentially leading to issues for galaxy surveys using UV-selected galaxies.
Our work demonstrates that the evolution of cluster galaxies continues well into their lifetime in the cluster and departs from a typical central galaxy evolutionary path.
\end{abstract}


\section{Introduction}
\label{sec:intro}

The environment of galaxy clusters has a significant impact on galaxy evolution compared to galaxies in the field.
Cluster galaxies tend to be quenched since their gas is stripped as they enter the dense cluster environment.
This causes them to be older, redder and more elliptical than their counterparts in the field \citep[e.g.][and references therein]{Dressler1980,Cooper2006,Donnari2021}.
The appearance and current state of a galaxy can vary significantly depending on its environment and how that environment interacts with the internal physical processes within the galaxy.

The exact processes involved in the transition between field galaxies and clusters are still under debate.
Ram pressure stripping from the intra-cluster medium pushes gas out of galaxies as they move through the cluster, with the exact details of such stripping dependent on the orbital parameters and intrinsic properties of the galaxy, as well as those of the host cluster and surrounding environment \citep{Gunn1972,Abadi1999}.

Various galaxy formation simulations using large cosmological volumes have broadly reproduced a wide range of galaxy properties and scaling relations.
\citet{Pillepich2018b} showed that galaxies in IllustrisTNG follow a well-defined stellar-to-halo mass relation  comparable to those expected in observations, and \citet{Nelson2018} showed the distribution of colours and magnitudes matched the distribution from observational surveys.
Galaxies in the EAGLE simulations \citep{Schaye2015} also broadly reproduce stellar masses and star formation rates for galaxies across a large range of redshifts \citep{Furlong2015}.

However, these studies typically encompass the entire galaxy population to understand the statistical properties of galaxies.
While it is well known that galaxies change in different environments, e.g. clusters or field, there is not currently a systematic study of the evolution of scaling relations in these different environments.
Because much of our understanding relies on inferring galactic properties through observable signatures, a detailed model for how the environment affects these relationships is essential if we want to fully describe how galaxy evolution interacts with different environments.

In the cluster environment, several important galaxy properties undergo different evolution than in the field or for centrals.
The stellar mass function, for example, typically changes in a cluster \citep{Ahad2021}.
The mode of star formation also varies depending on the length of time spent in a cluster, and the presence and morphology of nearby galaxies can alter a galaxy's evolution \citep{Perez2023},

Cluster properties, like the baryon fraction within clusters, stabilises early on and remains relatively constant for much of the Universe's history \citep{Chiu2018}.
This makes clusters an interesting test for studying environmental impacts since the effect is fairly constant over time.
Thus, galaxies that have been in a cluster for an extended period of time experienced similar effects as recently accreted galaxies but for a longer duration.
Changes in the galaxy properties, therefore, can be attributed the same environmental effect rather than being confounded with an environment that changes with redshift.

Clusters are large structures in the cosmic web that is spread throughout the universe and made up of dark matter.
Because dark matter is invisible but dominates the gravitational force in the universe, galaxies are often used as tracers to understand the structure of the underlying dark matter \citep[e.g.][]{Gregory1978,Bond1996b,Colless2001,Sarkar2023}.
However, it is also known that galaxies are biased tracer of the matter distribution in a way that varies depending on galaxy sample and redshift \citep[for review, see][and references therein]{Desjacques2018}.

Of additional interest to this work is the use of galaxies to measure the splashback radius ($R_{\rm sp}$).
$R_{\rm sp}$ is a description of the boundary of a dark matter halo that is rooted in orbital dynamics and encloses the first orbits of infalling material \citep{Diemer2014, Adhikari2014, More2015}.
Unlike spherical overdensity definitions like $R_{\rm200}$, $R_{\rm sp}$ does not suffer from pseudo-evolution, the apparent growth of a halo due to the expansion of the universe rather than changes in the halo itself \citep{Diemer2013}.
The accumulation of first apocentres causes a caustic in the radial density profile of haloes \citep{Huss1999,Adhikari2014}.

This significant decrease in the gradient of the radial density profile, hereafter referred to as the ``splashback feature'', has been used to identify the radius where the steepest slope occurs ($R_{\rm st}$) in theoretical and observational studies \citep[e.g.][]{More2015,Deason2020,Xhakaj2020,O'Neil2021,Baxter2017,Shin2019}.
$R_{\rm st}$ is often used as a proxy for $R_{\rm sp}$ since, in spherical collapse scenarios, infalling shells will pile up near the apocentre of their orbits and create a caustic \citep{Bertschinger1985}.
It is important to note that although the point of steepest slope and the radius enclosing the first orbit of infalling material are closely related quantities and are both often referred to as $R_{\rm sp}$, they do not correspond exactly with each other \citep{Diemer2020a}.
Although they are not identical quantities, $R_{\rm st}$ is significantly easier to measure both observationally and in simulations and is thus used in a wide range of studies \citep[e.g.][]{More2015,Baxter2017,Deason2020,O'Neil2021,O'Neil2022}.
In this work, we also focus on identifying the splashback feature and use the point of steepest slope to calculate $R_{\rm st}$.

Galaxy number density profiles of clusters show a similar splashback feature as is found in the dark matter density profile \citep[e.g.][]{Baxter2017,O'Neil2021}.
This makes using number densities of galaxies in clusters a promising technique to measure $R_{\rm st}$ in observations.
Although promising, the splashback feature in galaxy profiles does not exactly align with the dark matter splashback radius \citep{Deason2020,O'Neil2021,O'Neil2022}.
The difference between $R_{\rm st}$ measurements in galaxy populations and in dark matter density profiles depends on several factors, such as the mass or star formation rate of the galaxies.

These differences may stem from the different evolutionary histories of the galaxy population used to trace out the splashback feature.
The more recently that galaxies have fallen into a cluster, the smaller the measured $R_{\rm st}$ is predicted to be \citep{Adhikari2021}.
This measurement refers to the steepest point in the radial density profile and not the orbital track of the infalling haloes which, while related, do not correspond exactly.
Importantly, the difference between these two measurements increases if the selected galaxies do not accurately represent the gravitational potential of the cluster.

The relationship between galaxy and dark matter distributions is useful both for calibrating observations of $R_{\rm st}$ and for understanding a cluster's influence on galaxy formation.
Identifying the galaxies that trace the dark matter splashback feature can lead to insight into the galaxy's response to the baryons in the cluster \citep[and more]{Bullock2002}.
Additionally, the quenching timescale evolves with time as the universe changes, so galaxies that fall into clusters earlier than others quench differently \citep{Hough2023}.
This makes it particularly important to understand how galaxies change within their environment so we can disentangle how galaxy evolution changes through cosmic history.

In this paper, we explore the relationship between residence time, i.e. the time since a galaxy first crossed $R_{\rm200,mean}$ of its host group, and other galactic properties.
We also investigate the relationship between galaxy residence time and $R_{\rm st}$, the steepest point in the radial density profile.
This helps establish $R_{\rm st}$ as a potential observable signature of the time a population of galaxies has spent within a cluster and whether residence time may be a consistent factor in $R_{\rm st}$ bias for different populations of galaxies.
The paper is organised as follows.
In Section \ref{sec:methods}, we describe the simulations (\ref{sec:methods:simulations}), cluster and galaxy samples (\ref{sec:methods:samples}), and analysis methods for calculating residence time (\ref{sec:methods:residence_time}) and $R_{\rm st}$ (\ref{sec:methods:splashback}).
We describe our results for how galaxy properties and scaling relations change in Sections \ref{sec:results_residence} and \ref{sec:results_scaling}, and the dynamics of these galaxies and how this affects $R_{\rm st}$ in Sections \ref{sec:results_dynamics} and \ref{sec:results_profiles}.
Finally, we summarise our findings in Section \ref{sec:conclusions}.

\section{Methods}
\label{sec:methods}

In this paper, we analyse how galaxy properties change with the residence time of galaxies in their host haloes using the IllustrisTNG simulations.
In this section, we describe the simulations, how properties are extracted from the simulations, and our method of calculating the splashback radius using subsets of galaxies and clusters.

\subsection{Simulations}
\label{sec:methods:simulations}

In this work, we study the residence time of galaxies within clusters in the IllustrisTNG simulations \citep{Nelson2018, Marinacci2018, Springel2018, Naiman2018, Pillepich2018b}.
The cosmological parameters used are
$\Omega_{\rm m} = \Omega_{\rm{dm}}+\Omega_{\rm{b}} = 0.3089,\ \Omega_{\rm{b}} = 0.0486,\ \Omega_{\Lambda} = 0.6911,\ \sigma_8=0.8159,\ n_s=0.9667$, and Hubble constant $H_0=100h\,\rm{km} \,\rm{s}^{-1}\,\rm{Mpc}^{-1}$ where $h=0.6774$ consistent with \cite{PlanckCollaborationXIII2016}.

These simulations use \textsc{Arepo} \citep{Springel2010,Weinberger2020}, a magnetohydrodynamic moving-mesh code, and a galaxy formation updated from the Illustris project \citep{Vogelsberger2014b}.
This includes a supernova wind model \citep{Pillepich2018a}, radio mode active galactic nuclei feedback \citep{Weinberger2017}, and updated numerical schemes \citep{Pakmor2016}.
Gas cells cool radiatively, including metal line cooling, and stochastically form stars through a two-phase effective equation of state \citep{Springel2003} that then return mass and energy to the gas through winds and supernova explosions.
Black holes are seeded in haloes following \citet{DiMatteo2005} and grow through gas accretion and mergers.
Feedback is injected into the gas depending on the accretion rate \citep{Weinberger2017}.
This produces a range of galaxy types and realistic clusters \citep[e.g.][]{Vogelsberger2014a, Vogelsberger2018, Barnes2018a, Genal2018,  Donnari2021}.

We use the highest resolution level of the largest box in the simulation suite with full baryonic physics, referred to as TNG300-1.
The box has periodic boundary conditions with a side length of $302\,\rm{Mpc}$.
There are $2\times2500^3$ cells/particles, with a target gas cell mass of $1.1\times10^7\,\rm{M}_{\odot}$ and a dark matter particle mass of $5.9\times10^7\,\rm{M}_{\odot}$.
The gravitational softening length of the dark matter particles is $1.5\,\rm{kpc}$ in physical (comoving) units for $z\leq1$ $(z>1)$.
The gas cells use an adaptive comoving softening that reaches a minimum of $0.37\,\mathrm{kpc}$.

\subsection{Galaxy and cluster samples}
\label{sec:methods:samples}

Haloes are identified within the simulation using a Friends-of-Friends (FoF) algorithm \citep{Davis1985}, which groups particles based on their spatial distribution.
Subhaloes are identified using the \textsc{Subfind} algorithm \citep{Springel2001,Dolag2009}, which identifies gravitationally bound particles.
The most bound subhalo within a group is considered the main halo.

The haloes are traced through snapshots using the {\textsc{LHaloTree}} algorithm, which links subhaloes based on the particles found in subsequent snapshots, and then the haloes are linked according to the most bound subhaloes \citep{Springel2005b}.

Our galaxies are hosted by subhaloes with a gravitationally bound mass greater than $10^{9}$ M$_{\odot}$.
We use the associated stars to calculate galaxy properties like stellar mass and star formation rate.

We use the same sample of galaxies and clusters as in \citet{Borrow2023}, which is similar to the selection used in the previous splashback radius studies of \citet{O'Neil2021} and \citet{O'Neil2022} with an additional criterion. We direct readers to those prior works for a full description of the selection, but briefly discuss it here for completeness.

Our selected host haloes have a mass $M_{\rm200,mean}$ between $10^{13}$ M$_{\odot}$ and $10^{15}$ M$_{\odot}$ at redshift $z=0$.
There are three haloes with mass greater than $10^{15}$~M$_\odot$ in the simulation, which we do not include since this is not enough haloes to form larger mass bins, i.e. $10^{15-15.5}$~M$_\odot$.
We remove haloes that fall within 10 $R_{\rm200,mean}$ of a more massive halo to mitigate effects from more massive haloes disrupting the outer regions of less massive haloes. 

Due to the requirement that the central clusters must be well-tracked through the merger trees, we removed several clusters (of varying masses) whose first progenitor tracks were either too short to accurately track in-fall over the required 10 Gyr period, or had discontinuities in their co-moving position tracks. This filtering was performed programatically at first and then confirmed with visual inspection of the tracks, leaving us with 1302 clusters in the sample.

Finally, we group these cluster haloes into four mass ranges: $10^{13-13.5}$ M$_{\odot}$, $10^{13.5-14}$ M$_{\odot}$, $10^{14-14.5}$ M$_{\odot}$, and $10^{14.5-15}$ M$_{\odot}$ that are used throughout the remainder of the paper.

\subsection{Galaxy residence time}
\label{sec:methods:residence_time}

All subhaloes begin the simulation with a residence time of zero.
The first progenitors of all of the clusters are tracked back the simulation from $z=0$ as far as possible (typically to around $z=15$), with their co-moving position and $R_{\rm 200, mean}(t)$ recorded at each snapshot. As noted above, at this stage some clusters are rejected due to an inability to track the first progenitor well.

The second stage of residence time calculation is to find all subhaloes within 10 $R_{\rm 200, mean}$ of the cluster (including other centrals), and track them in the same way. Once the list of positions has been created, the radial distance between the center of each subhalo and the cluster, $r(t)$ is calculated at each snapshot. The first (physical) time at which the galaxy falls within $R_{\rm 200, mean}$ (i.e. when $r(t) < R_{\rm 200, mean}(t)$ for the smallest value of $t$) is recorded as the `infall time', $t_{\rm infall}$ of the galaxy. The residence time is then calculated as $t_{\rm res} = t_{\rm Hubble} - t_{\rm infall}$, i.e. the difference between the age of the universe at $z=0$ and when the galaxy first entered $R_{\rm 200, mean}$ of the cluster progenitor.

We must include all galaxies within a wide search radius around the $z=0$ cluster due to the presence of backsplash galaxies, which would be missed if we restricted our search to only galaxies `resident' in the cluster (measured as those with $r < R_{\rm 200, mean}$) at $z=0$. \citet{Borrow2023} found, using the exact same cluster selection, a significant population of galaxies at $r(z=0) > R_{\rm 200, mean}$ that had previously been within the sphere of influence of the galaxy cluster.
With the splashback radius of our cluster sample typically being $R_{\rm st} \approx 1.2-1.4R_{\rm 200, mean}$, using the steepest slope as a proxy for $R_{\rm sp}$ and calculated as described in Section \ref{sec:methods:splashback}, we further expect a significant population of galaxies that have had close encounters with the central cluster galaxy that now have large orbital radii.

A drawback to our residence time calculation procedure is that it is only calculated at snapshot output times, and no interpolation is employed. This means the temporal fidelity of our calculation is reduced slightly, and means that our residence times are a strict lower limit. In this study, however, we restrict our residence time binning to 1 Gyr wide bins, and with a maximal relevant snapshot spacing of 0.25 Gyr in IllustrisTNG, we find in practice that this is not a significant issue.

\subsection{The splashback radius of clusters}
\label{sec:methods:splashback}

In this work, we measure the splashback radius $R_{\rm sp}$ of a stacked set of clusters following the methods in \citet{O'Neil2022}.
We use the point of steepest slope in the stacked radial density profile as a proxy for $R_{\rm sp}$ since theoretically there is a caustic in idealised scenarios where the first apocentres of orbits occur.
This differs in practice from the exact definition of the splashback radius \citep{More2015,Diemer2020a}, and we differentiate these values by referring to the point of steepest slope as $R_{\rm st}$ and the splashback radius as $R_{\rm sp}$.

For each host halo in our sample, we calculate the radial number density of galaxies from the halo centre.
We calculate the galaxy number density within 32 log-spaced bins between $0.1-5\ R_{\rm200,mean}$ of the halo.
To calculate $R_{\rm sp}$ for a given galaxy residence time range, we use only galaxies that fall within this range to construct the number density profile.

As described in Section \ref{sec:methods:samples}, we divide our host haloes into four mass ranges.
For each host halo and galaxy residence time range, we stack the density profiles using the mean density in each bin.
We bootstrap the sample used to create the stacked profile 32 times with replacement, which gives an error bar on each point in the profile.
We then fit the functional form from \citet{Diemer2014}.
This function combines the Einasto profile to describe the inner regions of a halo that transitions to an outer region that flattens out the mean density of the universe $\rho_{\rm mean}$:
\begin{align}
\label{eq:density}
    \indent\rho(r) &= \rho_{\rm{inner}} \times f_{\rm{trans}} + \rho_{\rm{outer}} \\
    \indent\rho_{\rm{inner}} &= \rho_{\rm{Einasto}} = \rho_{\rm{s}} \exp{\left( -\frac{2}{\alpha} \left[\left(\frac{r}{r_{\rm{s}}}\right)^{\alpha} - 1 \right] \right)}  \nonumber \\
    \indent f_{\rm{trans}} &=  \left[ 1 + \left(\frac{r}{r_{\rm{t}}}\right)^\beta \right]^{-\frac{\gamma}{\beta}} \nonumber \\
    \indent\rho_{\rm{outer}} &= \rho_{\rm mean} \left[b_e \left( \frac{r}{5R_{\rm200,mean}} \right)^{-S_e} +1 \right]\:. \nonumber
\end{align}
The parameters $\rho_s, r_s, r_t, \alpha, \beta, \gamma, b_e,$ and $S_e$ are left free to vary for a given fit.

Following past work \citep{O'Neil2021,O'Neil2022}, we fit the logarithmic slope to the derivative of Equation \ref{eq:density} since the feature we are interested in is more significant in the profile gradient than in the profile.
When the profile is fit, we identify the minimum of the gradient as $R_{\rm sp}$.
We bootstrap the haloes included in the stacked profile 1024 times to obtain an error on our $R_{\rm sp}$ measurement, in addition to the 32 bootstraps used to obtain each $R_{\rm sp}$ sample.
When fitting profiles that have a very low central density gradient (>-1.5, as in the case using galaxies with residence times between 0 and 1 Gyrs), we exclude the first several bins from our fitting so the profile resembles the functional form given in Equation \ref{eq:density} and it is possible to identify the steepening slope near the edge of the halo.

\section{Results}
\label{sec:results}

In this section, we discuss how galaxy populations differ depending on how long they have lived within their host halo.

\subsection{Properties of galaxies and residence times} 
\label{sec:results_residence}

We first discuss how common galaxy properties and scaling relations change depending on the residence time of galaxies.
We separate the galaxies around host haloes by the time since they first fell past $R_{\rm200,mean}$ and explore the differences between them.

\begin{figure*}
    \centering
    \includegraphics{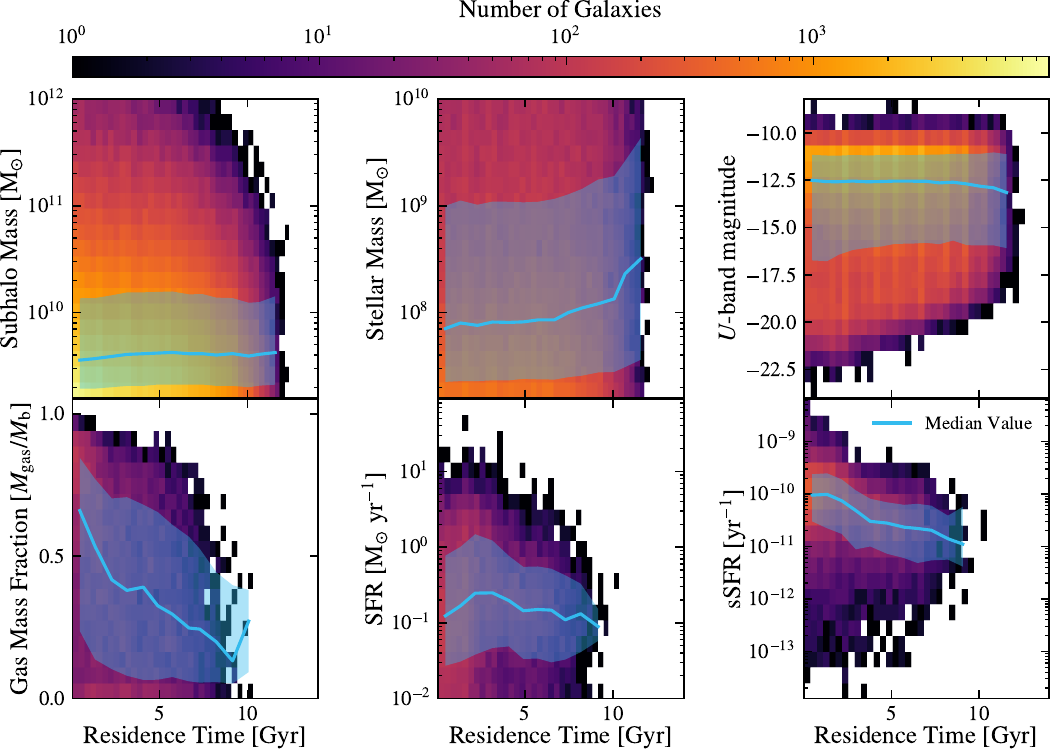}
    \caption{Each panel shows a 2D histogram of a given subhalo property and subhalo residence time in its host halo.  We do not include haloes with a residence time of zero, as these are galaxies that have not yet interacted with a host cluster halo. In each panel we also show a running median to demonstrate the overall trend of the population as well as the 16-84 percentile range. The galaxies included in this figure are only those that survive to $z=0$, excluding central haloes, not all possible galaxies within the entire history of the simulation. The sSFR, SFR, and gas fraction do not include galaxies with a zero value within the running median, which is the majority of galaxies (as shown by the background colour map).}
    \label{fig:properties_restime}
\end{figure*}

We show in Figure \ref{fig:properties_restime} the distribution of several galaxy properties as a function of galaxy residence time.
These plots exclude the central galaxy and show only surrounding haloes.
All properties use values within twice the 3D stellar half-mass radius of the subhaloes for consistent comparison with each other and to be more similar to observations, except for the total subhalo mass which is the total bound mass.

In the top left panel, we show the total gravitationally bound subhalo mass.
There are a number of features immediately clear in this panel. There is an increasing number of subhaloes present with decreasing residence time, and the median subhalo mass is relatively flat with residence time up to around one cluster crossing time $t_{\rm cross} \approx 5$ Gyr, after which it begins to decrease.
This is due to a number of factors that have been studied in prior work. 

First, as the clusters themselves grow in mass over time, they gain access to both more abundant and more massive subhaloes as the subhalo mass function evolves \citep[see e.g.][]{Giocoli2008}.
Subhaloes that are resident in the cluster are also subject to tidal stripping, where dark matter is removed from their outskirts over time. This will decrease the bound dark matter mass of subhaloes.
The final plausible explanation for the decrease in median subhalo mass as a function of residence time is dynamical friction, which will ensure that the most massive subhaloes merge with the central galaxy on timescales similar to $t_{\rm cross} =5$ Gyr, with all of these cumulative effects known as `segregation' \citep{vandenBosch2016}.

The average stellar mass within twice the half-mass radius (top middle) increases for higher residence times.
This indicates that galaxies with a high residence time consist of stellar populations with less luminosity per stellar mass, i.e. that the stars within the galaxies are less luminous than stars in more recently accreted galaxies.
The larger the residence time of a galaxy, the older its stellar population tends to be.
Galaxies with a residence time greater than 12 Gyr have stellar populations that are roughly 9 Gyr, while galaxies with a residence time of less than 1 Gyr have a stellar population that is aged about 1.5 Gyr.
Older stars are less luminous in the $U$-band, so the galaxies with a higher residence time have less luminous stellar populations.

In addition, the longer a galaxy has lived within the cluster, the more likely it is to have merged, which increases the stellar mass while the dark matter does not increase as significantly since it is more susceptible to stripping.
Thus, the high residence time galaxies have large populations of stars with lower $U$-band luminosity.
It is also possible that we are missing low-stellar mass galaxies that are below our resolution limit.
Galaxies that have been in the cluster for a long time and been stripped or disrupted would then no longer be identified in the simulation, contributing to the increase in average stellar mass.

In the lower left, we show the gas fraction within twice the half-mass radius for galaxies, which is calculated by dividing the total gas mass by the total baryonic mass.
This quantity sharply declines with residence time as gas is quickly stripped as galaxies fall into clusters and groups.
It then continues to decrease slowly as star formation depletes the remaining gas reserves.

The star formation rate (lower middle) correspondingly decreases with residence time.
The star formation rate shown here is the average star formation rate of star particles over 1 Gyr.
We note that we show only galaxies with nonzero star formation rates.
Previous studies have found that the star formation rate has some dependence on distance from the cluster centre \citep{Oyarzun2023}.
Since galaxies that are farther out tend to be recently accreted galaxies \citep{vandenBosch2016}, this is consistent with their proposed explanation that quenching is driven by environmental interactions, but it may also be due to the difference in the type of galaxies that were accreted at different times.
\citet{Finn2023} compared the star formation rate of field, infalling, and galaxies in the central regions of the cluster and found that field galaxies had higher star formation rates than infalling and cluster galaxies, but that the infalling and cluster populations did not significantly differ.
Other works, e.g. \citet{Paccagnella2016}, do find a significant difference between core cluster galaxies and infalling galaxies.
Our measured star formation rate drops very quickly and $\sim74$ percent of cluster galaxies have zero star formation rate across all residence times.

The specific star formation rate (lower right), which is the same star formation rate as the middle panel divided by the stellar mass of the galaxies, has a flatter trend.
The quenched fraction of galaxies, where a galaxy is quenched if the sSFR is less than $10^{-11}\ \rm{yr}^{-1}$ and includes galaxies with zero star formation, remains above 99 percent for all residence times greater than zero and increases with residence time.
Thus, even though the stellar mass increases, the calculated SFR excludes the majority of galaxies and decreases.

\subsection{Impact of residence time on galaxy scaling relations}
\label{sec:results_scaling}

Previous studies have investigated the evolution of scaling relations with redshift, and the star formation rate increases with redshift in both observations \citep[e.g.][]{Whitaker2014,Speagle2014} and in simulations \citep[e.g.][]{Donnari2019}, with \citet{Speagle2014} showing that this trend continues to redshift $z\approx6$.
This finding describes star-forming galaxies, which are primarily those outside a cluster.
Once the galaxy enters the cluster, it becomes quenched and is no longer considered when developing these relations.
Thus, these scaling relations with redshift are a good predictor of the conditions of galaxies at infall but do not correspond with the scaling relations as a function of residence time. 
Since these relations describe star-forming galaxies, they do not necessarily describe the different population of quenched cluster galaxies.
These scaling relations are well-established, but their usefulness is limited when the galaxies of interest differ from those used to construct the relations.
In our work, 96 percent of galaxies have a residence time of 0.

\begin{figure*}
    \centering
    \includegraphics[width=.95\linewidth]{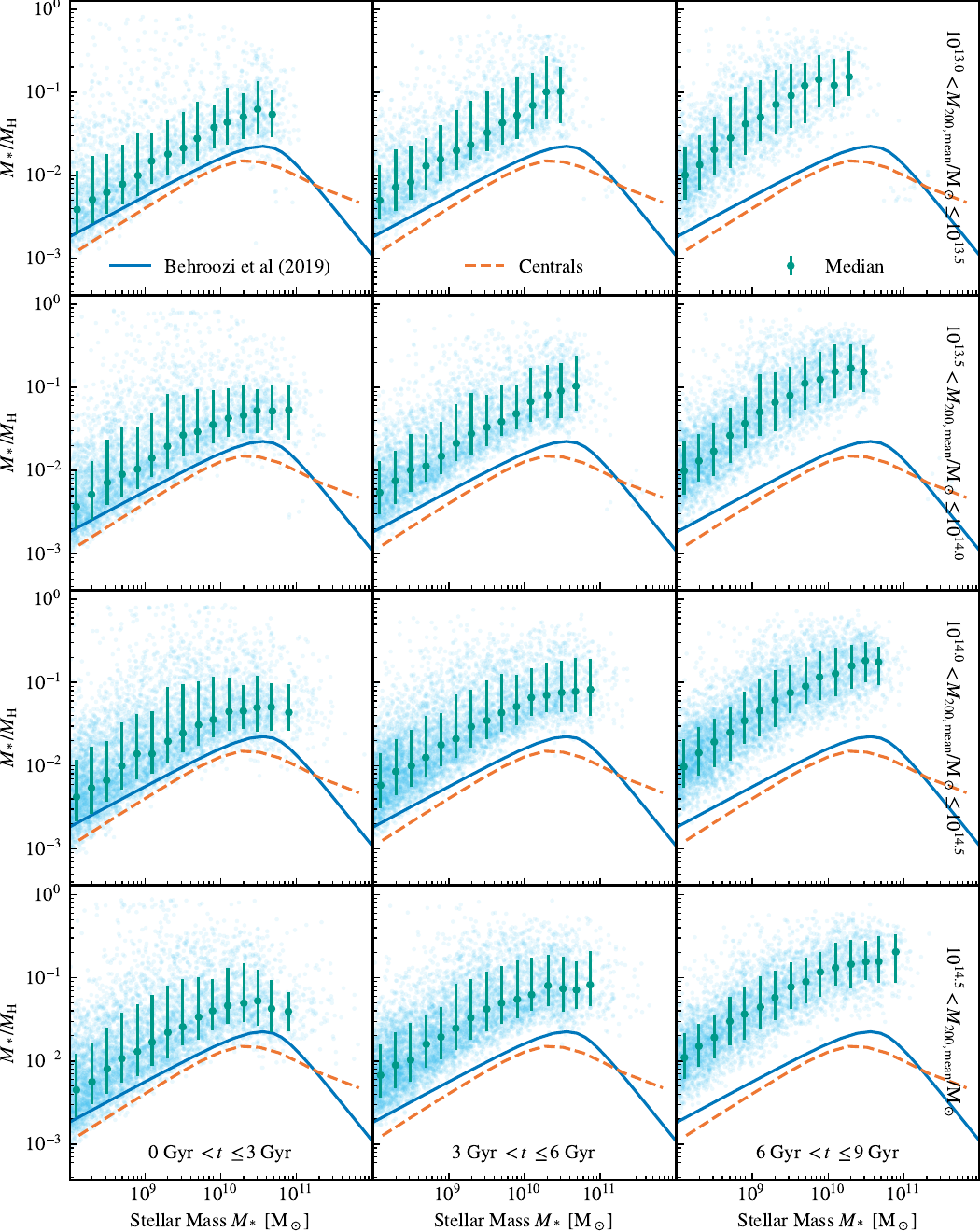}
    \caption{The stellar-to-halo mass ratio as a function of stellar mass for three bins in residence time (columns), split by cluster mass (rows). We show the trend of the median for all centrals as the orange line for reference, alongside the abundance matching results from \citet{Behroozi2019}. For each panel we show the median in green, 16-84 percentile range as error-bars, and entire population as the background scatter.  We see that as residence time increases to the right, the galaxies move further away from the median central and \citet{Behroozi2019} relations.  This shift occurs for all cluster mass ranges in similar quantities.}
    \label{fig:smhmrresidence}
\end{figure*}

Figure \ref{fig:smhmrresidence} shows the stellar-to-halo mass ratio as a function of residence time, where $M_\star$ is the galaxy stellar mass within twice the half-mass radius, and $M_{\rm H}$ is the total bound subhalo mass.
The points in the left panel shows the relationship for galaxies that have a residence time of less than 3 Gyr, not including galaxies that have zero residence time, i.e. galaxies that have never been within $R_{\rm200,mean}$ of a host halo.
The middle panel shows the relationship for galaxies with a residence time between 3 and 6 Gyr, and the right panel shows galaxies between 6 and 9 Gyr.
Each light blue point represents one galaxy within the residence time range and the teal points show the median at that mass with the error bars showing the 16$^{\rm th}$ and 84$^{\rm th}$ percentiles.
The dashed orange line shows the median value for all central galaxies within the TNG300 simulation for reference, and the solid blue line shows the relationship from \citet{Behroozi2019}, both of which are the same in every panel.
As the residence time increases, from the left to right panels, the galaxies move further away from the \citet{Behroozi2019} relationship to a higher stellar-to-halo mass ratio. The offest from the central relation is stable with host cluster mass.

\citet{Wang2018} found, using the New York University Value Added Galaxy Catalog from the SDSS Data Release 7, along with supplemental information from various sources, that satellite galaxies, at a fixed halo mass, have a stellar mass that is typically 0.5 dex higher than their field counterparts. That is broadly consistent with what we find here (in the central column of panels with residence times $t_{\rm res} \approx 5$~Gyr), though we do find significantly higher offsets for galaxies that have resided in the cluster for a longer time.

\citet{Rodriguez2021} also found a significant difference between satellite and central galaxies in the scaling relations of galaxies in SDSS and TNG300, in particular the stellar-to-halo mass ratio and galaxy size.
The difference between central and satellite galaxies in their study was much larger for the galaxy size than the stellar-to-halo mass relation.
This hints at the different way that centrals and satellites grow within the cluster, mainly that central galaxies can grow significantly through accretion and mergers while satellites are also affected by processes like stripping that minimise halo growth.

In semi-analytical galaxy formation modeling using the {\textsc{mpld2-sag}} model, \citet{Ruiz2023} found that cluster galaxies can lose approximately 50 percent of their halo mass over time as they interact with a cluster host, a value broadly similar to the offset that we see between our $0<t/{\rm Gyr}\leq3$ and $6<t/{\rm Gyr}\leq9$ panels.

This also highlights the difference between the evolution of cluster galaxies and the overall evolution of galaxies through cosmic time.
The longer that a galaxy lives within a cluster, the higher the stellar-to-halo mass ratio tends to be.
This has been noted previously \citep[e.g.][]{Knebe2011}, and is due to the dark matter halo being more easily stripped than the more central stellar halo and galaxy \citep[e.g.][]{Fattahi2018}.
The main point here is that this greatly influences the expected scaling relation depending on how long the galaxy population has been influenced by a cluster environment.
Other studies \citep[e.g.][]{vandenBosch2008} note only a mild dependence of scaling relations on galaxy environment and that environmental effects are primarily driven by the stellar mass of galaxies.
They argue that galaxies of similar stellar masses that reside in different environments have similar scaling relations, and that the environmental dependence stems from certain stellar masses being more abundant in certain environments.
Here we show that the scaling relations retain strong dependence on stellar mass but there is also significant variation in the scaling relations with galaxy residence time even for galaxies of similar stellar masses.
We attribute this change primarily to the differing residence times of the galaxies and not due to these populations having different spatial distributions, which is discussed further in Appendix \ref{ap:radius}.

\begin{figure}
    \centering
    \includegraphics{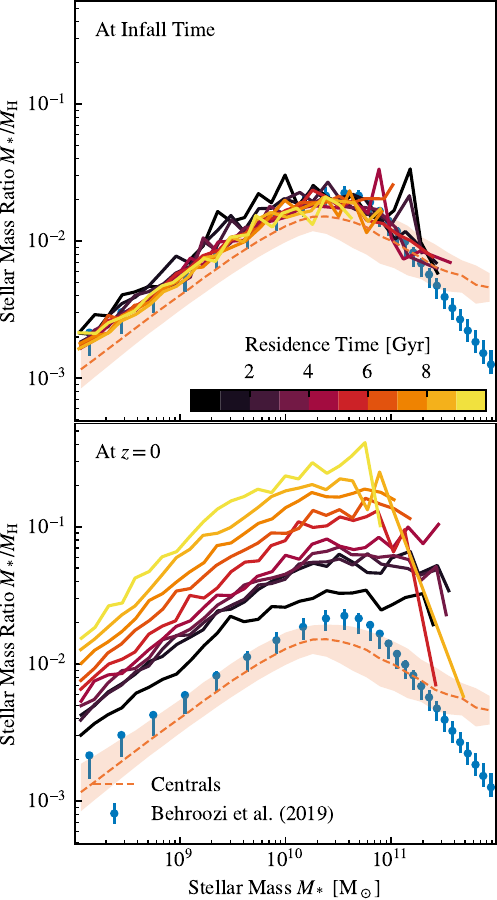}
    \caption{\emph{Both panels}: The blue points and errorbars show comparative abundance-matching data from \citet{Behroozi2019}, representing the ratio between galaxy stellar mass and host halo mass. The orange line shows the median trend of all central galaxies in TNG300, with the orange shaded region representing the 16-84 percentile range. \emph{Top panel}: Each line represents the median of a subset of the $z=0$ cluster galaxies tracked back to their infall time. \emph{Bottom panel}: Shows the same galaxies (i.e. each line colour in the top and bottom panel represents the exact same selection of unique substructures) but now showing their properties at $z=0$.}
    \label{fig:mass_smhm}
\end{figure}

\begin{figure}
    \centering
    \includegraphics{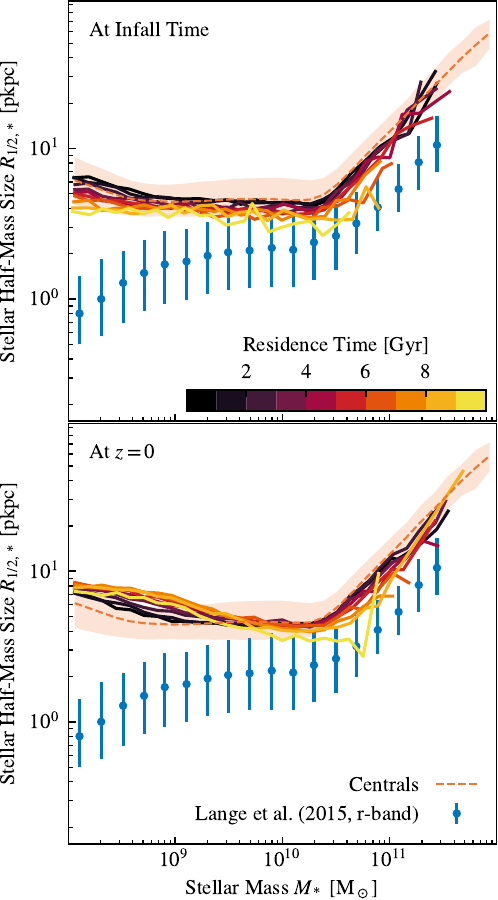}
    \caption{\emph{Both panels}: The blue points and errorbars show observational data from \citet{Lange2015}, representing the two dimensional half-mass radius of galaxies as a function of their mass. The orange line shows the median trend (three-dimensional aperture) of all central galaxies in TNG300, with the orange shaded region representing the 16-84 percentile range. \emph{Top panel}: Each line represents the median of a subset of the $z=0$ cluster galaxy tracked back to their infall time. \emph{Bottom panel}: Shows the same galaxies (i.e. each line colour in the top and bottom panel represents the exact same selection of unique substructures) but now showing their properties at $z=0$.}
    \label{fig:mass_radius}
\end{figure}

\begin{figure}
    \centering
    \includegraphics{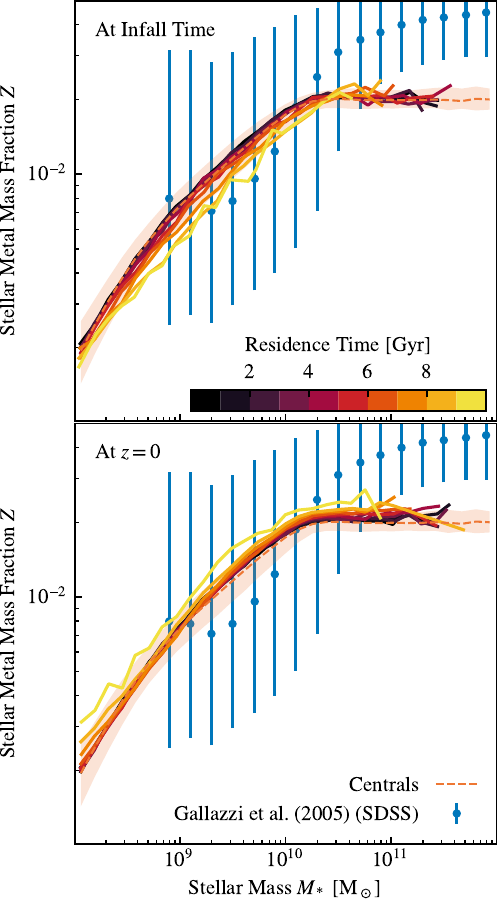}
    \caption{\emph{Both panels}: The blue points and errorbars show comparative observational data from \citet{Gallazzi2005}, showing the galaxy stellar metal mass fraction (corrected from solar relative metallicity in the cited text). The orange line shows the median trend of all central galaxies in TNG300, with the orange shaded region representing the 16-84 percentile range. \emph{Top panel}: Each line represents the median of a subset of the $z=0$ cluster galaxy tracked back to their infall time. \emph{Bottom panel}: Shows the same galaxies (i.e. each line colour in the top and bottom panel represents the exact same selection of unique substructures) but now showing their properties at $z=0$.}
    \label{fig:mass_metallicity}
\end{figure}

We expand our study of scaling relations in Figures \ref{fig:mass_smhm}-\ref{fig:mass_metallicity}.
Each figure here shows a different galaxy scaling relation, with the differently coloured lines representing the median value for galaxies of different residence times (analogously to the teal points in Figure \ref{fig:smhmrresidence}). We show all the scaling relations as a function of the galaxy stellar mass contained within twice the stellar half-mass radius for consistency.
All galaxies represented within the figure are those with residence times greater than zero at $z=0$ for the entire cluster sample. In the top panel of each figure, we track all of these galaxies back to their infall time and show their properties at that time (i.e. one $t_{\rm res}$ ago). In the bottom panel, we show the properties of the galaxies at $z=0$ to highlight how the galaxies have evolved whilst inside the cluster environment.
Finally, the dashed lines show the relationship for all centrals in the volume at $z=0$, with a shaded region representing the 16$^{\rm th}$-84$^{\rm th}$ percentile range, and the blue points show data from other, typically observational, sources for reference.

Figure \ref{fig:mass_smhm} shows the stellar-to-halo mass ratio as in Figure \ref{fig:smhmrresidence}.
We show, for comparison, the relationship from \citet{Behroozi2019} in the blue points.
The relations for the galaxies at infall time are similar and the differences between the populations occur at $z=0$, with all clear, systematic, evolution seen in their properties after cluster-driven evolution.
This indicates that galaxies are similar to each other and the known scaling relations prior to being influenced by the cluster environment, and the stellar-to-halo mass ratio increases the longer galaxies are in the cluster.

The stellar-to-halo mass ratio increases with residence time as the dark matter is stripped more readily than the stellar mass.
This occurs for galaxies of all masses and the shape of this relation remains relatively constant and is shifted higher.
\citet{Bahe2017} similarly found a difference in the stellar-to-halo mass relation between non-centrals in the Hydrangea simulations and non-centrals in observations, although they consider cluster-mass haloes rather than galaxy-mass haloes as we do.
We note again here that the galaxies included in the upper panel are the same galaxies as in the lower panel, just tracked back to their own infall time.

We do see a significantly higher median stellar mass ratio for a fixed stellar mass at infall time than the centrals.
This implies that the efficiency of galaxy formation is higher in the dense environment surrounding the galaxy cluster.
Pre-processing can also lower the total mass of galaxies, which would also tend to increase the stellar mass ratio \citep[e.g.][]{Lopes2024}.
Although there are some centrals included in our sample of infalling galaxies, the two samples have distinct underlying influences on their evolution trajectory.
Our sample consists of galaxies that are near large clusters and subject to dense environments and pre-processing while central galaxies in general include a large number of field galaxies.

\citet{vanderBurg2020} found, using the GOGREEN survey using deep Gemini/GMOS spectroscopy, a significant offset in the abundance of high stellar mass galaxies within the cluster environment compared to the field at $z\approx1$. They found that there was a factor $\approx 2$ higher abundance of galaxies with $M_\star \approx 10^{10.5}$~M$_\odot$ in the high density environment compared to the field on average, which corresponds to our factor of $\approx 0.2$ dex offset in stellar mass ratio at a fixed stellar mass. They found in addition, however, that there was an \emph{lower} abundance of galaxies with $M_\star < 10^{9.5}$~M$_\odot$ in the cluster environment, which would appear to be in tension with our results from TNG300. A full study of the impact of environment on the abundances of galaxies, however, is out of the scope of this paper.

Using galaxies in high-density environments at $2.0 \leq z \leq 4.2$ in the ZFOURGE survey, \citet{Hartzenberg2023} found significant evolution in the mean stellar mass of galaxies, with higher density environments consistently hosting higher-mass galaxies (by a factor of $\approx 20$ percent, consistent with our findings here). These galaxies were also significantly more likely to be quiescent (by a factor $\approx 2-3$) than those in intermediate-density environments, supporting the idea that these galaxies have their stellar mass `locked in' and then the ratio $M_\star / M_{\rm H}$ evolves due to a reduction in halo mass.

Figure \ref{fig:mass_smhm} highlights that differences in the cluster galaxy population are not due to older galaxies being `stalled' (as they are typically quenched upon entering the cluster) and retaining a high stellar-to-halo mass ratio, but rather that the galaxies, or their haloes, undergo significant evolution within the cluster environment.
\citet{Hough2023} find, using a semi-analytic model and simulated galaxy clusters, that a non-negligible fraction of galaxies need several passes through a cluster before they are completely quenched.
\citet{Ruiz2023} find, using simulated galaxies in the \textsc{MDPL2-SAG} catalogue, that most galaxies make at least one passage through a cluster before being stripped of dark matter and gas.
For this stripping to occur, a galaxy must pass near the centre of the cluster.
This indicates continued stripping and galaxy evolution as the galaxies fall in, leading to the trend we see here, where galaxies continue to change long after they have entered a host halo.
\citet{Donnari2021} also notes that satellites accreted more than approximately 5 Gyr ago are almost entirely passive but that more recently accreted galaxies are not, indicating a long time scale for stripping and quenching to occur.

Figure \ref{fig:mass_radius} shows the physical radius enclosing half the stellar mass of a galaxy as a function of galaxy stellar mass with points taken from \citet{Lange2015} for reference. There is a consistent offset between the TNG300 data and the observations that can be attributed to the use of a 3D aperture in TNG and a 2D aperture in the observations of \citet{Lange2015}. The global average sizes of galaxies predicted by the TNG model have been shown previously to be consistent with observations \citep{Pillepich2019}.

At infall time, galaxies with a mass $M_\star \lesssim 10^{10}$~M$_\odot$ have monotonically decreasing sizes as a function of residence time. This is mainly caused by softening-driven inflation of these galaxies, with the physical softening length being smaller at higher redshift. This softening inflation effect has been studied previously outside of TNG \citep[e.g.][]{Ludlow2019} and with the TNG model in \citet{Pillepich2019}, with sizes in the TNG model having a floor at roughly three times the gravitational softening \citep[see also][]{Power2003}. As galaxies evolve to lower redshift, their sizes increase in accordance with the increasing physical softening length.

At the lowest masses, galaxy sizes also increase as time is spent in the cluster. This is likely due to a number of both internal and external factors, as discussed in depth in Section 5 of \citet{Borrow2023}. Even though the galaxy sizes at masses $M_\star < 10^{9.5}$~M$_\odot$ are not converged within TNG300, \citet{Borrow2023} showed that those that have had significant interactions with galaxy clusters have inflated sizes (as shown directly here) due to tidal heating. Here, we show that galaxies that have been in the cluster the longest have seen the highest level of size inflation, even though they began with the smallest physical sizes when entering the cluster.

At the highest masses, we see significantly less size evolution with residence time. At these high masses, internal numerical processes such as dynamical heating \citep{Ludlow2023}, and external processes driven by cluster interaction, have been shown to have reduced impacts \citep{Borrow2023}, with the galaxy sizes merely `locked in' through an intrinsically high quiescent fraction. Galaxies with masses $M_\star > 10^{9.5}$~M$_\odot$ show little-to-no size evolution off the main track with stellar mass, though a small subset of galaxies within these tracks clearly show some significant star formation since initially entering the galaxy cluster, and alongside this they show size growth.

The median size for galaxies with $M_\star > 10^8$~M$_\odot$ within clusters is lower than those of $z=0$ centrals, with galaxies with a residence time of approximately 5 Gyr showing around a 50 percent reduction in physical size relative to a central at the same mass. Our finding that galaxies residing within clusters are smaller on average than centrals corresponds nicely to recent observational findings from \citet{Mosleh2020}, who showed using the CANDELS and 3D HST Legacy Program data that quiescent galaxies have, on average, around 0.5 dex smaller sizes than star-forming counterparts at $0.3 < z < 0.7$ at $M_\star \approx 10^{10}$ M$_\odot$. The majority ($75$ percent) of our cluster galaxy sample is classified as quiescent when making a cut on instantaneous star formation rate (generally due to having a gas fraction consistent with zero, and particularly with little cold gas).

Our galaxies also show, consistent with \citet{Mosleh2020}, a pivot mass leading to increasing galaxy sizes of $M_\star \approx 10^{10.5}$~M$_\odot$.
The trend of increasing galaxy sizes beyond the pivot mass of $M_\star > 10^{10.5}$~M$_\odot$ was demonstrated specifically within the environment of cluster Abell 209 by \citet{Annunziatella2016}, who found that passive galaxies within the cluster increase their size by approximately $1$ dex over $1$ dex increase in stellar mass, consistent with our results. We note that we refrain from comparing numerical values of sizes directly here due to our use of different tracers (mass and light), as well as inconsistencies due to aperture variations \citep{deGraaff2022}.
We see this pivot point across all residence times, so this feature is not related to processes within the cluster environment, such as stripping.
There is a strong correlation between stellar mass and the sSFR, with less massive galaxies having a larger sSFR.
Above $M_\star\approx10^{10.5}$~M$_{\odot}$, the slope of the relation steepens.
This indicates a difference in star forming and growth mechanisms.

In Figure \ref{fig:mass_metallicity}, we now focus on the stellar metallicities of the galaxies. Stellar metal mass fractions are calculated as the total metal mass in stellar particles within twice the half-mass radius, divided by the total stellar mass within that same aperture. This inevitably leads to potential offsets with residence time as we have already shown that the half-mass size sees significant evolution as galaxies reside in the cluster (Figure \ref{fig:mass_radius}). We compare again to all redshift $z=0$ central galaxies (orange dashed line), and observational data from the Sloan Digital Sky Survey \citep[SDSS;][]{Gallazzi2005}.

Galaxies with a longer residence time (i.e. those that fell in at higher redshift) have a lower median stellar metallicity at a fixed mass at infall time. This is an inevitable consequence of the increasing metal richness of the interstellar medium as galaxies are allowed to evolve to lower redshift, as shown in the TNG model in \citet{Nelson2018} and \citet{Torrey2019}. Galaxies residing in the cluster tend to have their cold gas stripped and depleted, with the stellar metallicity at infall hence corresponding roughly to the metallicity of that gas before any such stripping occurred.

In the lower panel, we show the $z=0$ relations binned by residence time. This panel provides significantly more puzzling results, with galaxies that have resided in the cluster for longer showing \emph{increasing} metallicity relative to those that have been resident for shorter times. Some of this effect, at least for galaxies with $M_\star > 10^9$~M$_\odot$, can be attributed to the smaller aperture that is used for older galaxies, leading to a preferential selection of the typically more metal rich galactic core \citep[see e.g.][for recent work]{Nanni2023}.
In tests of this, smaller apertures led to an increased metallicity.
With a metal rich galactic core, tidal stripping will also preferentially remove metal poorer stars, further increasing the metallicity of long-residence time galaxies.
In related theoretical work, \citet{Ruiz2023} found using a semi-analytic galaxy formation model, that the bulge fraction of cluster galaxies increases over time for a given stellar mass as they interact with galaxy clusters. 
This supports the trends we see here.

\begin{figure}
    \centering
    
    \includegraphics[width=\linewidth]{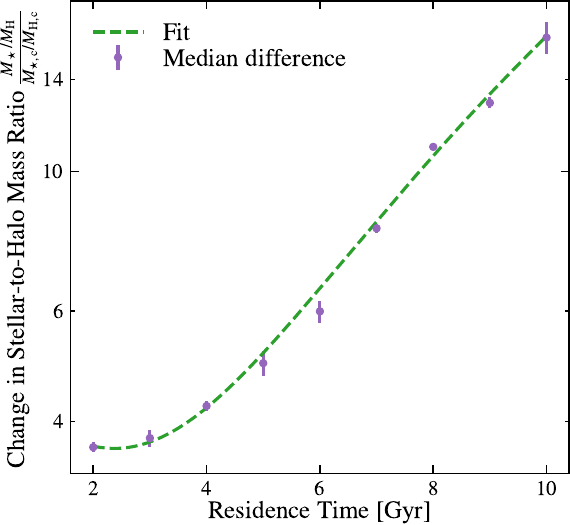}
    \caption{Purple points show the median difference in stellar-to-halo mass ratio between cluster galaxies of a given residence time and centrals ($M_{\rm c}$) at $z=0$ in TNG300.
    For each residence time, we calculate the median difference between galaxies at each stellar mass with the corresponding centrals, with the errorbars representing the 16-84 percentile range. The green dashed line shows the best fit line to a quadratic function. }
    \label{fig:fits}
\end{figure}

In Figure \ref{fig:fits}, we explicitly show how the stellar-to-halo mass fraction (Figure \ref{fig:mass_smhm}) changes with residence time.
The increase is fairly consistent across the stellar mass range.
For each residence time, we divide the stellar-to-halo mass ratio by the corresponding value for central galaxies at $z=0$ for a given stellar mass then take the median value of all stellar masses.
We use the 16th and 84th percentiles of galaxies at each residence time to estimate errors.
The evolution of these scaling relations is approximately quadratic, and we fit the functional form:
\begin{equation}
    X = a t_{\rm res}^2 + b t_{\rm res} + c
    \label{eq:quad_fit}
\end{equation}
where $X$ is the stellar-to-halo mass fraction, and $t_{\rm res}$ is the residence time.
The constants $a,b,$ and $c$ are fit parameters.
The scaling relation for centrals is then approximately $Xt_{\rm res}$.
For the stellar-to-halo mass ratio using residence times between 2 and 10 Gyr, we obtain values $a=0.220,\ b=-1.044$ and $c=4.865$ with a reduced $\chi^2=2.56$.

We do not use this to fit the stellar half-mass size since this value has significant mass dependence as well as residence time dependence.
In addition, while the difference in stellar metal mass fraction also appears approximately quadratic as a function of residence time, the errors are too large to provide an adequate fit and this increase is not understood as well.

\begin{figure*}
    \centering
    
    \begin{subfigure}{0.33\linewidth}
    \includegraphics[width=\linewidth]{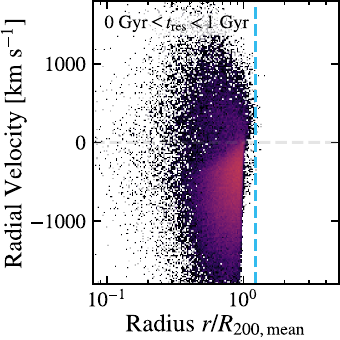}
    \end{subfigure}
    \begin{subfigure}{0.33\linewidth}
    \includegraphics[width=\linewidth]{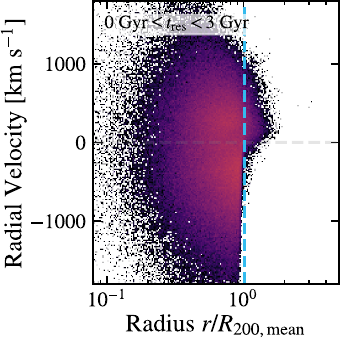}
    \end{subfigure}
    \begin{subfigure}{0.33\linewidth}
    \includegraphics[width=\linewidth]{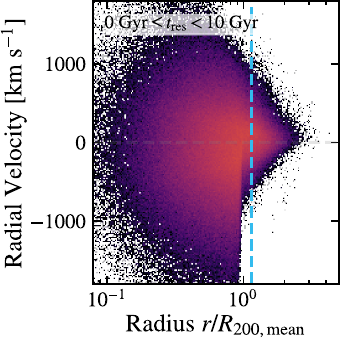}
    \end{subfigure}
    \vspace{10pt}
    
    \begin{subfigure}{0.33\linewidth}
    \includegraphics{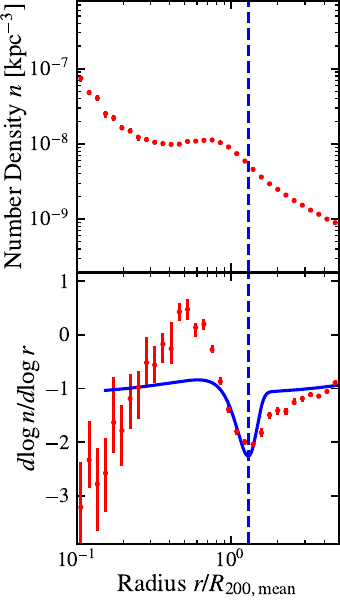}
    \end{subfigure}
    \begin{subfigure}{0.33\linewidth}
    \includegraphics{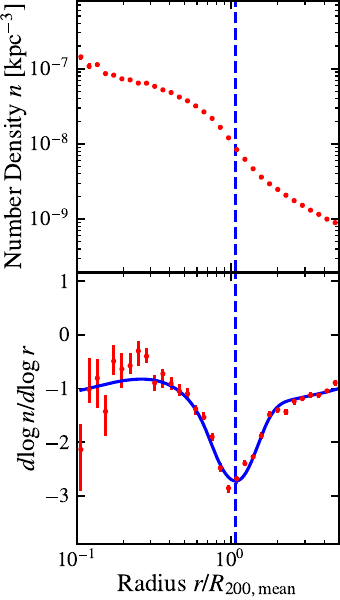}
    \end{subfigure}
    \begin{subfigure}{0.33\linewidth}
    \includegraphics{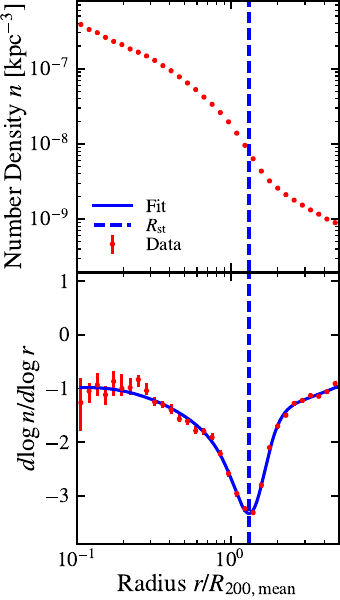}
    \end{subfigure}
    
    \caption{\textit{Top:} The phase space for galaxies with a residence time between 0-1 Gyr (left), subhaloes with a residence time between 0-3 Gyr (middle), or subhaloes with a residence time between 0-10 Gyr (right). The vertical dashed line shows the $R_{\rm st}$ calculated from bootstrapping the fitted differential density profile 1024 times and taking the median value.  As the maximum residence time increases, more subhaloes move away from the center of their host halo with a positive radial velocity.  More galaxies also reach larger radii when the residence time is increased.
    \textit{Middle, Bottom:} Example density profiles and gradients for all haloes with $M>10^{13}\ \rm{M}_{\odot}$ and subhaloes with a residence time between the same values as above.  Data points calculated from the simulation are shown in red, and the fit of the gradient is shown with the solid blue line.  $R_{\rm st}$ is the vertical dashed line as in the top panels.}
    \label{fig:halo_profile}
\end{figure*}

\subsection{Cluster density profiles}
\label{sec:results_profiles}

We now discuss how the dynamics of galaxies affects the measured density profiles of clusters.
We calculate the number density of galaxies as a function of radius for stacked sets of clusters and find the point of steepest slope $R_{\rm st}$ as discussed in Section \ref{sec:methods:splashback}.

In Figure \ref{fig:halo_profile}, we show three example profiles fitted using the procedure described in Section \ref{sec:methods:splashback}.
We stack the profiles for haloes with mass $M_{\rm200,mean}>10^{13}\: \rm{M}_{\odot}$.
In the left panel, the density profile is constructed using only galaxies with a residence less than 1 Gyr.
Galaxies with a residence time less than 3 Gyr are used to construct the density profile in the middle panel, and galaxies with a residence time less than 10 Gyr are used in the right panel.
We also show the radial velocity and distance from the host halo centre in the top panels.
Note that we show only galaxies that have entered the cluster in the phase diagram to understand the dynamics, but we must include galaxies beyond $R_{\rm200,mean}$ to obtain a density profile.

We use a maximum galaxy residence time rather than binned residence time since we must include galaxies in the outer region of the host halo.
Since galaxies that have never been within $R_{\rm200,mean}$ of a cluster have zero residence time and $R_{\rm sp}$ is typically larger than $R_{\rm200,mean}$, excluding these galaxies would remove the splashback feature in the profiles.

The series of panels demonstrates the conformance of infalling galaxies to the gravitational potential of the host halo.
In the left panel, the inner region of the density profile drops off very quickly since there are few galaxies that have recently accreted in that region of the clusters.
This also causes the density gradient to drop at low radii and makes the splashback feature less distinct. By construction, the splashback feature should not exist here yet; the galaxies have not had time to `splash back'. What the steepest slope measures here is the transition from the central density profile of the cluster to the more uniform density environment outside of the cluster.
In order to identify the feature with our fitting procedure, we exclude the bins before the peak in the gradient (e.g. near $0.7\:R_{\rm200,mean}$ in the left panel) as mentioned in Section \ref{sec:methods:splashback}.
The change in shape of the density profile around the splashback feature increases the discrepancy between the point of steepest slope and the true splashback radius of dark matter.
Therefore, the galaxy sample used to calculate the point of steepest slope must be carefully chosen for $R_{\rm st}$ to be used as an adequate proxy for $R_{\rm sp}$.

We can also see from the phase space of these subhaloes that the longer a galaxy has been inside a cluster, excluding galaxies with zero residence time, the farther away from the centre it can get.
For galaxies with a residence time less than 2 Gyr, most galaxies are still falling towards the centre of the halo.

\begin{figure}
    \centering
    \includegraphics{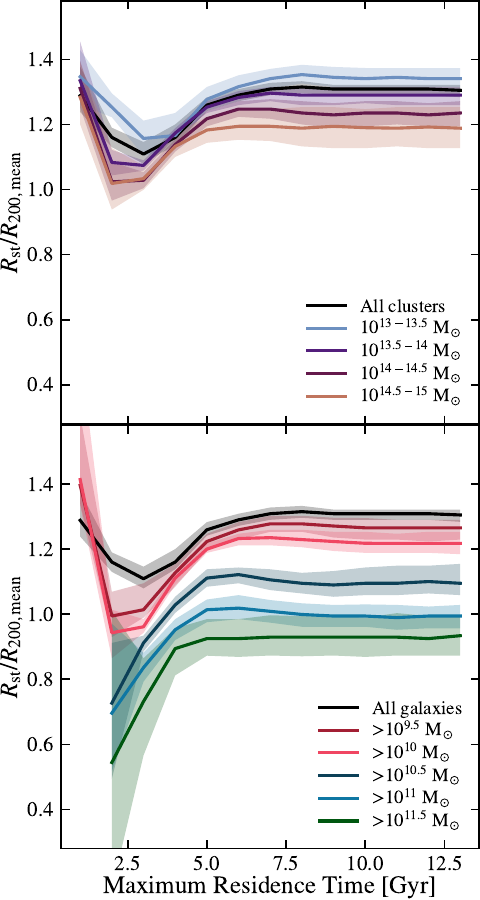}
    \caption{Each line shows the trend of $R_{\rm st}$ for a galaxy profile using galaxies with residence time up to the displayed value. The top panel shows all galaxies in host haloes within a given mass range, and the bottom shows galaxies above a given mass within all host haloes. Each line decreases at low residence time where data exists $t_{\rm res}\lesssim2.5$ before reaching a minimum and increasing until 5 Gyr, where they then level off.  Galaxies of all masses show similar behaviour, indicating that this effect influences galaxies and clusters similarly across mass scales.  However, there is still a strong mass dependence for both galaxy and host halo mass.}
    \label{fig:rsp_restime}

\end{figure}

Figure \ref{fig:rsp_restime} shows the evolution of the point of steepest slope as a function of maximal galaxy residence time (i.e. the evaluated points) include all galaxies with $t_{\rm res} \leq t_{\rm res, max}$.
The average residence time of the galaxies included in each profile is significantly less than the maximum residence time but increases steadily, and changing the figure to use the mean residence time rather than the maximum does not change our results.
In the top panel, each point shows the splashback radius using all galaxies in the stacked set of clusters in the given mass ranges from $10^{13-15}\ {\rm M}_{\odot}$.
In the bottom panel, the the galaxy number density of host haloes using galaxies above a minimum total subhalo mass between $10^{9.5-11.5}\ \rm{M}_{\odot}$ are stacked for all host haloes  with mass $M_{\rm200,mean}>10^{13}\ \rm{M}_{\odot}$.

Consistent with past work \citep[e.g.][]{More2015,Diemer2017b,O'Neil2021}, $R_{\rm st}/R_{\rm200,mean}$ decreases with host halo mass.
The $10^{14.5-15}\ \rm{M}_{\odot}$ host haloes consistently have the lowest $R_{\rm st}/R_{\rm200,mean}$.
$R_{\rm st}/R_{\rm200,mean}$ also decreases with total bound subhalo mass as expected \citep{O'Neil2022}.

The other distinct feature is that $R_{\rm st}/R_{\rm200,mean}$ initially decreases between residence times from 0-2 Gyr before increasing again.
Constructing the steepest slope and using it as a proxy for splashback radius is nonsensical for $t_{\rm res} > t_{\rm cross}$, as galaxies have not had enough time to reach pericentre and return to their apocentre, forming the splashback feature within the profile. 
The maximum residence time of 3 Gyr gives us the smallest value of $R_{\rm st}$.
They may have reached the pericentre, but most galaxies have not reached the apocentre of their orbit.
After 3 Gyr, more galaxies reach larger radii and $R_{\rm st}$ increases.
By 5 Gyr, many galaxies have had time to reach their apocentre and our measurements of $R_{\rm st}$ stabilise.
By this point, the galaxies have largely had enough time resident in the cluster to be representative of the internal dynamics.
\citet{Adhikari2021} predicted that $R_{\rm sp}/R_{\rm200,mean}$ increased with increasing residence times, and we are qualitatively consistent with their results for similar residence times.
Due to the constraints of their study, they used only three residence time bins and did not investigate residence times less than 1 Gyr, where we find the most significant decrease in $R_{\rm st}$.

Of additional interest in these plots is that these trends persist across residence times.
The effect of the cluster on the galaxy dynamics is constant across mass scales for both host haloes and galaxies.
However, there remains a strong dependence on mass even when the galaxies are split by residence time.
\citet{vandenBosch2016} and \citet{Joshi2017} noted that cluster galaxies are naturally segregated.
Galaxies that accreted earlier tend to be less massive and bound in tighter orbits since they have undergone more tidal stripping than galaxies accreted more recently.
We would therefore expect that the mass and resident times affect $R_{\rm st}$ in intertwining ways, but we see that these effects are fairly distinct in our plots.

Separating galaxies by residence time does not remove dependence on galaxy or host halo mass and vice versa.
This deviates from the predictions of \citet{O'Neil2022}, which suggested that differences in $R_{\rm st}$ for different galaxy populations was driven by whether the galaxy population had enough time to properly virialize with the host halo.
However, galaxies remain separated by mass, as suggested by \citet{vandenBosch2016}.
Host halo mass also appears to be fundamental to setting the splashback radius, contrary to \citet{More2015} but which was hinted at in \citet{Diemer2020}.
Early work studying $R_{\rm sp}$ attributed most mass evolution in the value of $R_{\rm sp}$ to differences in accretion rate, where higher accretion rates tend to push $R_{\rm sp}$ to smaller values, but more recent work that is able to more carefully separate accretion rate and mass find that binning by accretion rate does not completely remove halo mass dependence.

\begin{figure*}
    \centering
    \includegraphics{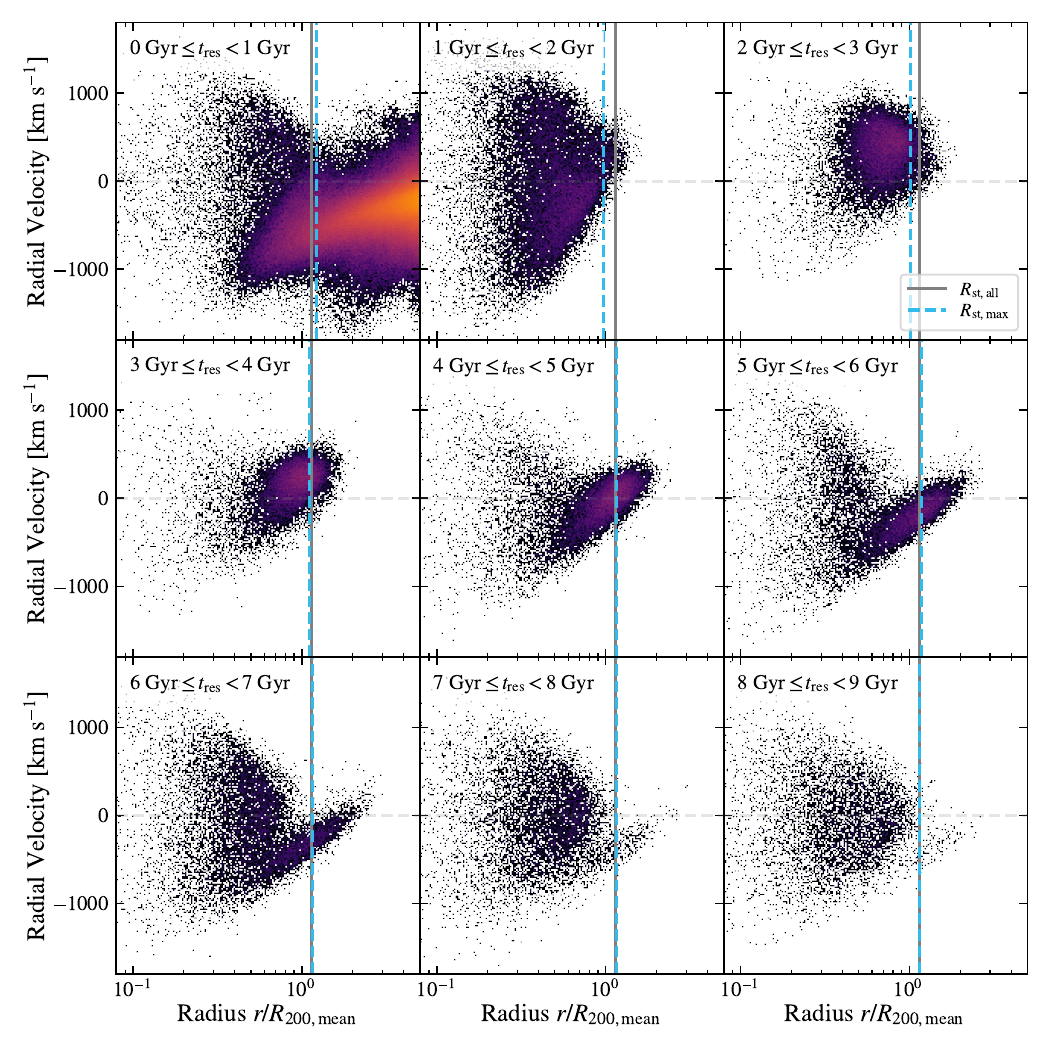}
    \caption{The phase space of galaxies in clusters between $10^{14-14.5}\ \rm{M}_{\odot}$ clusters split by residence time of galaxies.  The results do not change significantly for different host halo mass ranges.  In the first panel, we show galaxies with residence time of less than 1 Gyr and increase the range by 1 Gyr in each successive panel.  These are galaxies that live within their host haloes at $z=0$ and not tracing a single galaxy population as it moves through the host.  Differently to Fig.~\ref{fig:halo_profile}, we do not include galaxies that have never fallen within $R_{\rm200,mean}$ of their host halo in these sub-plots. The vertical grey lines show $R_{\rm st}$ calculated using all galaxies, and the vertical blue bashed lines show $R_{\rm st}$ using galaxies with residence times less than the upper limit given in each panel. We see that the galaxies have primarily negative velocity and large radius early on.  By 2 Gyr, a few galaxies have started to turn around, and most galaxies have reached this point by 3 Gyr.  By the crossing time, 5 Gyr, galaxies have had time to reach the edge of the host halo again.  As the galaxies fall back in, they disperse through the main halo.}
    \label{fig:phase}
\end{figure*}

\subsection{The evolution of galaxy dynamics}
\label{sec:results_dynamics}

We now discuss how galaxies differ dynamically for populations of different residence times.
We focus in particular on the movement of galaxies within clusters and how this affects the shapes of their number density profiles and the point of steepest slope in the profile.
The point of steepest slope $R_{\rm st}$ is a promising signature of the splashback radius $R_{\rm sp}$ but, as noted in previous work, is greatly impacted by the galaxy population used to make this measurement \citep{Adhikari2021,O'Neil2022}.

Figure \ref{fig:phase} shows the evolution of the phase space of galaxies with different residence times for clusters with mass $10^{14}\ \rm{M}_{\odot}<M<10^{14.5}\ \rm{M}_{\odot}$.
This sample of clusters is representative and the results do no change significantly for different host halo mass ranges.
We calculate the radial velocity and distance of the galaxies relative to the host halo centre, and we show all galaxies at $z=0$ around the selected clusters.
As discussed in Section \ref{sec:results_profiles}, we must use those with a residence time of $0$ Gyrs or less (i.e. those not resident in the cluster) for our fitting procedure in order to capture the splashback feature in the outskirts of the halo.

The top left panel shows galaxies that have a residence time less than 1 Gyr, including galaxies with zero residence time.
These galaxies primarily have a negative radial velocity, especially those with a radius less than $R_{\rm200,mean}$, indicating that they are falling towards the centre of the cluster.
They have also not been within $R_{\rm200,mean}$ for very long, if at all, so there are very few galaxies that reach low radii.
As we increase the residence time, galaxies have time to fall further into the cluster.
Galaxies with a residence time of 2-3 Gyr have had time to reach the pericentre of their orbit and start moving away from the cluster centre and predominantly have positive radial velocities.
At around 5 Gyr, we see that galaxies are crossing back out of the cluster and turning back in.

We include backsplash galaxies, i.e. galaxies that are outside the host halo at a given time but were previously inside the halo, since they share more similarities with cluster galaxies than those that have never been within a cluster \citep{Borrow2023}.
\citet{Diemer2017a} compared the dynamical splashback radius using particle trajectories to the point of steepest slope and found that $R_{\rm st}$ typically contains about 85 percent of particle apocenters.
Although this number is approximate and can vary and is affected by factors like mass and accretion rate, this provides an estimate for how many galaxies we expect to have splashed out of the host halo.
During a galaxy's backsplash period, it will not be subject to the same environmental factors like continual stripping, although it will be subject to others like depleted gas reservoirs.
This causes certain properties, like the gas fraction, to evolve differently than others, like the star formation rate, and these properties will behave more similarly to galaxies with a lower residence time.
Given that this population is approximately 15 percent of our sample of cluster galaxies, we expect this to primarily impact the scatter in our relations while the median remains fairly steady.

The evolution of the boundary of these galaxies corresponds to the typical crossing time, using the approximation $t_{\rm cross}=t_{\rm H}/(5\sqrt{\Omega_{\rm m}})$ \citep{Diemer2017b}.
For Hubble time $t_{\rm H}=1/(H_0)=14.4$ Gyr and $\Omega_{\rm m}=0.3$, we get a crossing time of 5.2 Gyr.
These galaxies have had time to reach the apocentre of their
orbit and are split between positive and negative radial velocities as the population turns around.
The splashback radius for the galaxies in a cluster should correspond to this apocentre, and we see that the measured steepest slope for galaxies with a residence time $\gtrsim 5$ Gyr aligns with the steepest slope using all galaxies at these residence times.
As the residence time of galaxies increases to 9 Gyr, the population of galaxies reaching their apocentre, visible as the distinct elliptical population in the previous panels, is less apparent.
These galaxies are integrated within the cluster, and they are spread more uniformly through the phase space within $R_{\rm st}$.

\section{Conclusions}
\label{sec:conclusions}

We used the largest simulation volume of the IllustrisTNG simulation suite, TNG300-1, to investigate the evolution of galaxies as they live in groups and clusters with mass $10^{13}<M/{\rm M}_{\odot}<10^{15}$.
We calculate the residence time of galaxies, defined as the time since they first fell within $R_{\rm200,mean}$ of their host halo, and examine the properties of these galaxies as a function of their resident time.
We summarise our findings as follows:
\begin{itemize}
    \item We show that there is evolution in several galaxy properties (subhalo mass, stellar mass, magnitude, gas mass fraction, SFR, and sSFR) in Figure \ref{fig:properties_restime}.  The median subhalo mass and magnitude slightly decrease with residence time while the median stellar mass slightly increases.  The gas mass fraction and SFR significantly decrease with resident time, but the sSFR decreases only slightly due to the increase in stellar mass.
    This is important because it emphasizes that galaxies continue their evolution within the cluster over large timescales up to 10 Gyr.
    \item We investigate the evolution of scaling relations in Figures \ref{fig:smhmrresidence}-\ref{fig:mass_metallicity}.  We show that for galaxies with longer residence times, the further their average value is from standard scaling relations.  The stellar-to-halo mass ratio and the stellar metallicity increase for galaxies at a given stellar mass, and the galaxy size and sSFR decrease.
    This highlights the effect of stripping and quenching that occurs within the cluster and how this significantly alters observable quantities.
    The cluster environment affects galaxies consistently across masses, and the scaling relations maintain their shape as the values are shifted up or down.  The evolution of the stellar-to-halo mass ratio with residence time is approximately quadratic with the form in Equation \ref{eq:quad_fit}. 
    \item Finally, we investigate how the point of steepest slope $R_{\rm st}$ in the galaxy number density profiles changes by using galaxies with different residence times to construct the density profile.  This value is often used as a proxy for the the splashback radius $R_{\rm sp}$ in galaxy clusters.  In Figure \ref{fig:rsp_restime}, we show that there is a strong dependence on residence time for residence times below 5 Gyr, which is approximately the time it takes for galaxies to cross the cluster.  For residence times less than 1 Gyr, $R_{\rm st}/R_{\rm200,mean}$ decreases.  Between 2 and 5 Gyr, $R_{\rm st}/R_{\rm200,mean}$ increases as predicted in previous studies.  It is therefore important to ensure that galaxies with residence times greater than 5 Gyr are included in the sample if the point of steepest slope is used to infer $R_{\rm sp}$.
    \item We also find that splitting galaxies by residence time does not remove the dependence of $R_{\rm sp}/R_{\rm200,mean}$ on host halo and galaxy mass, nor does splitting galaxies by host halo and galaxy mass remove the dependence on residence time.  We therefore conclude that all of these quantities are important to take into consideration when calibrating the point of steepest slope measurements to the value of the splashback radius.
\end{itemize}

Observationally, there are still many more complications that need to be accounted for.
Projection effects and foreground contamination, for example, likely smears these effects and makes them less distinct.
Here, we additionally note the influence of environment on observational signatures.

While it is well known that cluster galaxies differ on average from central and field galaxies, this paper emphasises that the relationship between galactic properties and between physical processes differ in a consistent and predictable way.
The role of the environment alters the evolutionary path of a galaxy such that our understanding and predictions for these galaxies must be adjusted.
Importantly, these differences do not occur immediately upon entering the cluster but continue to evolve for many Gyrs from the galaxies' infall.

\section{Acknowledgements}
We thank the anonymous referee for thorough and thoughtful comments that improved the draft.
We also thank Yannick Bah\'{e} for helpful discussion.
Some of the computations were performed on the Engaging cluster supported by the Massachusetts Institute of Technology. MV acknowledges support through NASA ATP 19-ATP19-0019, 19-ATP19-0020, 19-ATP19-0167, and NSF grants AST-1814053, AST-1814259, AST-1909831, AST-2007355 and AST-2107724.

We made use of the following software for the analysis:
\begin{itemize}
	\item {\textsc{Python}}: \citet{vanRossum1995}
	\item {\textsc{Matplotlib}}: \citet{Hunter2007}
	\item {\textsc{SciPy}}: \citet{Virtanen2020}
	\item {\textsc{NumPy}}: \citet{Harris2020}
	\item {\textsc{Astropy}}: \citet{Astropy2013,Astropy2018}
	\item {\textsc{SwiftSimIO}}: \citet{Borrow2020, Borrow2021}
\end{itemize}

\section{Data Availability}
The data is based on the IllustrisTNG simulations that are publicly available at \url{https://tng-project.org} \citep{Nelson2019}.
Reduced data are available upon request.

\bibliographystyle{mn2e}
\bibliography{bibliography}

\begin{thebibliography}{97}
\expandafter\ifx\csname natexlab\endcsname\relax\def\natexlab#1{#1}\fi

\bibitem[{{Abadi}, {Moore} \& {Bower}(1999){Abadi}, {Moore}, \&
  {Bower}}]{Abadi1999}
{Abadi} M.~G., {Moore} B., {Bower} R.~G., 1999, \mnras, 308, 947

\bibitem[{{Adhikari}, {Dalal} \& {Chamberlain}(2014){Adhikari}, {Dalal}, \&
  {Chamberlain}}]{Adhikari2014}
{Adhikari} A., {Dalal} N., {Chamberlain} R.~T., 2014, Jounral of Cosmology and
  Astroparticle Physics, 8

\bibitem[{{Adhikari} {et~al}\mbox{.}(2021){Adhikari}, {Shin}, {Jain}, {Hilton},
  {Baxter}, {Chang}, {Wechsler}, {Battaglia}, {Bond}, {Bocquet}, {Choi},
  {DeRose}, {Devlin}, {Dunkley}, {Evrard}, {Ferraro}, {Hill}, {Hughes},
  {Gallardo}, {Lokken}, {MacInnis}, {Madhavacheril}, {McMahon}, {Nati},
  {Newburgh}, {Niemack}, {Page}, {Palmese}, {Partridge}, {Rozo}, {Rykoff},
  {Salatino}, {Schillaci}, {Sehgal}, {Sif{\'o}n}, {To}, {Wollack}, {Wu}, {Xu},
  {Aguena}, {Allam}, {Amon}, {Annis}, {Avila}, {Bacon}, {Bertin}, {Bhargava},
  {Brooks}, {Burke}, {Rosell}, {Kind}, {Carretero}, {Castander}, {Choi},
  {Costanzi}, {da Costa}, {Vicente}, {Desai}, {Diehl}, {Doel}, {Everett},
  {Ferrero}, {Fert{\'e}}, {Flaugher}, {Fosalba}, {Frieman},
  {Garc{\'\i}a-Bellido}, {Gaztanaga}, {Gruen}, {Gruendl}, {Gschwend},
  {Gutierrez}, {Hartley}, {Hinton}, {Hollowood}, {Honscheid}, {James},
  {Jeltema}, {Kuehn}, {Kuropatkin}, {Lahav}, {Lima}, {Maia}, {Marshall},
  {Martini}, {Melchior}, {Menanteau}, {Miquel}, {Morgan}, {L.~C. Ogando},
  {Paz-Chinch{\'o}n}, {Malag{\'o}n}, {Sanchez}, {Santiago}, {Scarpine},
  {Serrano}, {Sevilla-Noarbe}, {Smith}, {Soares-Santos}, {Suchyta}, {E.~C.
  Swanson}, {Varga}, {Wilkinson}, {Zhang}, {Austermann}, {Beall}, {Becker},
  {Denison}, {Duff}, {Hilton}, {Hubmayr}, {Ullom}, {Lanen}, {Vale}, {Vale}, \&
  {Vale}}]{Adhikari2021}
{Adhikari} S. {et~al.}, 2021, \apj, 923, 37

\bibitem[{{Ahad} {et~al}\mbox{.}(2021){Ahad}, {Bah{\'e}}, {Hoekstra}, {van der
  Burg}, \& {Muzzin}}]{Ahad2021}
{Ahad} S.~L., {Bah{\'e}} Y.~M., {Hoekstra} H., {van der Burg} R. F.~J.,
  {Muzzin} A., 2021, \mnras, 504, 1999

\bibitem[{{Annunziatella} {et~al}\mbox{.}(2016){Annunziatella}, {Mercurio},
  {Biviano}, {Girardi}, {Nonino}, {Balestra}, {Rosati}, {Bartosch Caminha},
  {Brescia}, {Gobat}, {Grillo}, {Lombardi}, {Sartoris}, {De Lucia}, {Demarco},
  {Frye}, {Fritz}, {Moustakas}, {Scodeggio}, {Kuchner}, {Maier}, \&
  {Ziegler}}]{Annunziatella2016}
{Annunziatella} M. {et~al.}, 2016, \aap, 585, A160

\bibitem[{{Astropy Collaboration} {et~al}\mbox{.}(2018){Astropy Collaboration},
  {Price-Whelan}, {Sip{\H{o}}cz}, {G{\"u}nther}, {Lim}, {Crawford}, {Conseil},
  {Shupe}, {Craig}, {Dencheva}, {Ginsburg}, {VanderPlas}, {Bradley},
  {P{\'e}rez-Su{\'a}rez}, {de Val-Borro}, {Aldcroft}, {Cruz}, {Robitaille},
  {Tollerud}, {Ardelean}, {Babej}, {Bach}, {Bachetti}, {Bakanov}, {Bamford},
  {Barentsen}, {Barmby}, {Baumbach}, {Berry}, {Biscani}, {Boquien}, {Bostroem},
  {Bouma}, {Brammer}, {Bray}, {Breytenbach}, {Buddelmeijer}, {Burke},
  {Calderone}, {Cano Rodr{\'\i}guez}, {Cara}, {Cardoso}, {Cheedella}, {Copin},
  {Corrales}, {Crichton}, {D'Avella}, {Deil}, {Depagne}, {Dietrich}, {Donath},
  {Droettboom}, {Earl}, {Erben}, {Fabbro}, {Ferreira}, {Finethy}, {Fox},
  {Garrison}, {Gibbons}, {Goldstein}, {Gommers}, {Greco}, {Greenfield},
  {Groener}, {Grollier}, {Hagen}, {Hirst}, {Homeier}, {Horton}, {Hosseinzadeh},
  {Hu}, {Hunkeler}, {Ivezi{\'c}}, {Jain}, {Jenness}, {Kanarek}, {Kendrew},
  {Kern}, {Kerzendorf}, {Khvalko}, {King}, {Kirkby}, {Kulkarni}, {Kumar},
  {Lee}, {Lenz}, {Littlefair}, {Ma}, {Macleod}, {Mastropietro}, {McCully},
  {Montagnac}, {Morris}, {Mueller}, {Mumford}, {Muna}, {Murphy}, {Nelson},
  {Nguyen}, {Ninan}, {N{\"o}the}, {Ogaz}, {Oh}, {Parejko}, {Parley}, {Pascual},
  {Patil}, {Patil}, {Plunkett}, {Prochaska}, {Rastogi}, {Reddy Janga},
  {Sabater}, {Sakurikar}, {Seifert}, {Sherbert}, {Sherwood-Taylor}, {Shih},
  {Sick}, {Silbiger}, {Singanamalla}, {Singer}, {Sladen}, {Sooley},
  {Sornarajah}, {Streicher}, {Teuben}, {Thomas}, {Tremblay}, {Turner},
  {Terr{\'o}n}, {van Kerkwijk}, {de la Vega}, {Watkins}, {Weaver}, {Whitmore},
  {Woillez}, {Zabalza}, \& {Astropy Contributors}}]{Astropy2018}
{Astropy Collaboration} {et~al.}, 2018, \aj, 156, 123

\bibitem[{{Astropy Collaboration} {et~al}\mbox{.}(2013){Astropy Collaboration},
  {Robitaille}, {Tollerud}, {Greenfield}, {Droettboom}, {Bray}, {Aldcroft},
  {Davis}, {Ginsburg}, {Price-Whelan}, {Kerzendorf}, {Conley}, {Crighton},
  {Barbary}, {Muna}, {Ferguson}, {Grollier}, {Parikh}, {Nair}, {Unther},
  {Deil}, {Woillez}, {Conseil}, {Kramer}, {Turner}, {Singer}, {Fox}, {Weaver},
  {Zabalza}, {Edwards}, {Azalee Bostroem}, {Burke}, {Casey}, {Crawford},
  {Dencheva}, {Ely}, {Jenness}, {Labrie}, {Lim}, {Pierfederici}, {Pontzen},
  {Ptak}, {Refsdal}, {Servillat}, \& {Streicher}}]{Astropy2013}
{Astropy Collaboration} {et~al.}, 2013, \aap, 558, A33

\bibitem[{{Bah{\'e}} {et~al}\mbox{.}(2017){Bah{\'e}}, {Barnes}, {Dalla
  Vecchia}, {Kay}, {White}, {McCarthy}, {Schaye}, {Bower}, {Crain}, {Theuns},
  {Jenkins}, {McGee}, {Schaller}, {Thomas}, \& {Trayford}}]{Bahe2017}
{Bah{\'e}} Y.~M. {et~al.}, 2017, \mnras, 470, 4186

\bibitem[{{Barnes} {et~al}\mbox{.}(2018){Barnes}, {Vogelsberger}, {Kannan},
  {Marinacci}, {Weinberger}, {Springel}, {Torrey}, {Pillepich}, {Nelson},
  {Pakmor}, {Naiman}, {Hernquist}, \& {McDonald}}]{Barnes2018a}
{Barnes} D.~J. {et~al.}, 2018, \mnras, 481, 1809

\bibitem[{{Baxter} {et~al}\mbox{.}(2017){Baxter}, {Chang}, {Jain}, {Adhikari},
  {Dalal}, {Kravtsov}, {More}, {Rozo}, {Rykoff}, \& {Sheth}}]{Baxter2017}
{Baxter} E. {et~al.}, 2017, \apj, 841

\bibitem[{{Behroozi} {et~al}\mbox{.}(2019){Behroozi}, {Wechsler}, {Hearin}, \&
  {Conroy}}]{Behroozi2019}
{Behroozi} P., {Wechsler} R.~H., {Hearin} A.~P., {Conroy} C., 2019, \mnras,
  488, 3143

\bibitem[{{Bertschinger}(1985)}]{Bertschinger1985}
{Bertschinger} E., 1985, \apjs, 58, 39

\bibitem[{{Bond}, {Kofman} \& {Pogosyan}(1996){Bond}, {Kofman}, \&
  {Pogosyan}}]{Bond1996b}
{Bond} J.~R., {Kofman} L., {Pogosyan} D., 1996, \nat, 380, 603

\bibitem[{{Borrow} \& {Borrisov}(2020)}]{Borrow2020}
{Borrow} J., {Borrisov} A., 2020, The Journal of Open Source Software, 5, 2430

\bibitem[{{Borrow} \& {Kelly}(2021)}]{Borrow2021}
{Borrow} J., {Kelly} A.~J., 2021, arXiv e-prints, arXiv:2106.05281

\bibitem[{{Borrow} {et~al}\mbox{.}(2023){Borrow}, {Vogelsberger}, {O'Neil},
  {McDonald}, \& {Smith}}]{Borrow2023}
{Borrow} J., {Vogelsberger} M., {O'Neil} S., {McDonald} M.~A., {Smith} A.,
  2023, \mnras, 520, 649

\bibitem[{{Bullock}, {Wechsler} \& {Somerville}(2002){Bullock}, {Wechsler}, \&
  {Somerville}}]{Bullock2002}
{Bullock} J.~S., {Wechsler} R.~H., {Somerville} R.~S., 2002, \mnras, 329, 246

\bibitem[{{Chiu} {et~al}\mbox{.}(2018){Chiu}, {Mohr}, {McDonald}, {Bocquet},
  {Desai}, {Klein}, {Israel}, {Ashby}, {Stanford}, {Benson}, {Brodwin},
  {Abbott}, {Abdalla}, {Allam}, {Annis}, {Bayliss}, {Benoit-L{\'e}vy},
  {Bertin}, {Bleem}, {Brooks}, {Buckley-Geer}, {Bulbul}, {Capasso},
  {Carlstrom}, {Rosell}, {Carretero}, {Castander}, {Cunha}, {D'Andrea}, {da
  Costa}, {Davis}, {Diehl}, {Dietrich}, {Doel}, {Drlica-Wagner}, {Eifler},
  {Evrard}, {Flaugher}, {Garc{\'\i}a-Bellido}, {Garmire}, {Gaztanaga},
  {Gerdes}, {Gonzalez}, {Gruen}, {Gruendl}, {Gschwend}, {Gupta}, {Gutierrez},
  {Hlavacek-L}, {Honscheid}, {James}, {Jeltema}, {Kraft}, {Krause}, {Kuehn},
  {Kuhlmann}, {Kuropatkin}, {Lahav}, {Lima}, {Maia}, {Marshall}, {Melchior},
  {Menanteau}, {Miquel}, {Murray}, {Nord}, {Ogando}, {Plazas}, {Rapetti},
  {Reichardt}, {Romer}, {Roodman}, {Sanchez}, {Saro}, {Scarpine}, {Schindler},
  {Schubnell}, {Sharon}, {Smith}, {Smith}, {Soares-Santos}, {Sobreira},
  {Stalder}, {Stern}, {Strazzullo}, {Suchyta}, {Swanson}, {Tarle}, {Vikram},
  {Walker}, {Weller}, \& {Zhang}}]{Chiu2018}
{Chiu} I. {et~al.}, 2018, \mnras, 478, 3072

\bibitem[{{Colless} {et~al}\mbox{.}(2001){Colless}, {Dalton}, {Maddox},
  {Sutherland}, {Norberg}, {Cole}, {Bland-Hawthorn}, {Bridges}, {Cannon},
  {Collins}, {Couch}, {Cross}, {Deeley}, {De Propris}, {Driver}, {Efstathiou},
  {Ellis}, {Frenk}, {Glazebrook}, {Jackson}, {Lahav}, {Lewis}, {Lumsden},
  {Madgwick}, {Peacock}, {Peterson}, {Price}, {Seaborne}, \&
  {Taylor}}]{Colless2001}
{Colless} M. {et~al.}, 2001, \mnras, 328, 1039

\bibitem[{{Cooper} {et~al}\mbox{.}(2006){Cooper}, {Newman}, {Croton}, {Weiner},
  {Willmer}, {Gerke}, {Madgwick}, {Faber}, {Davis}, {Coil}, {Finkbeiner},
  {Guhathakurta}, \& {Koo}}]{Cooper2006}
{Cooper} M.~C. {et~al.}, 2006, \mnras, 370, 198

\bibitem[{{Davis} {et~al}\mbox{.}(1985){Davis}, {Efstathiou}, {Frenk}, \&
  {White}}]{Davis1985}
{Davis} M., {Efstathiou} G., {Frenk} C.~S., {White} S.~D.~M., 1985, \apj, 292,
  371

\bibitem[{{de Graaff} {et~al}\mbox{.}(2022){de Graaff}, {Trayford}, {Franx},
  {Schaller}, {Schaye}, \& {van der Wel}}]{deGraaff2022}
{de Graaff} A., {Trayford} J., {Franx} M., {Schaller} M., {Schaye} J., {van der
  Wel} A., 2022, \mnras, 511, 2544

\bibitem[{{Deason} {et~al}\mbox{.}(2020){Deason}, {Fattahi}, {Frenk}, {Grand},
  {Oman}, {Garrison-Kimmel}, {Simpson}, \& {Navarro}}]{Deason2020}
{Deason} A.~J., {Fattahi} A., {Frenk} C.~S., {Grand} R. J.~J., {Oman} K.~A.,
  {Garrison-Kimmel} S., {Simpson} C.~M., {Navarro} J.~F., 2020, \mnras, 496,
  3929

\bibitem[{{Desjacques}, {Jeong} \& {Schmidt}(2018){Desjacques}, {Jeong}, \&
  {Schmidt}}]{Desjacques2018}
{Desjacques} V., {Jeong} D., {Schmidt} F., 2018, \physrep, 733, 1

\bibitem[{{Di Matteo}, {Springel} \& {Hernquist}(2005){Di Matteo}, {Springel},
  \& {Hernquist}}]{DiMatteo2005}
{Di Matteo} T., {Springel} V., {Hernquist} L., 2005, \nat, 433, 604

\bibitem[{{Diemer}(2017)}]{Diemer2017a}
{Diemer} B., 2017, \apj, 20

\bibitem[{{Diemer}(2020{\natexlab{a}})}]{Diemer2020a}
{Diemer} B., 2020{\natexlab{a}}, \apjs, 251, 17

\bibitem[{{Diemer}(2020{\natexlab{b}})}]{Diemer2020}
{Diemer} B., 2020{\natexlab{b}}, \apj, 903, 87

\bibitem[{{Diemer} \& {Kravtsov}(2014)}]{Diemer2014}
{Diemer} B., {Kravtsov} A.~V., 2014, \apj, 789, 18

\bibitem[{{Diemer} {et~al}\mbox{.}(2017){Diemer}, {Mansfield}, {Kravtsov}, \&
  {More}}]{Diemer2017b}
{Diemer} B., {Mansfield} P., {Kravtsov} A.~V., {More} S., 2017, \apj, 843, 140

\bibitem[{{Diemer}, {More} \& {Kravtsov}(2013){Diemer}, {More}, \&
  {Kravtsov}}]{Diemer2013}
{Diemer} B., {More} S., {Kravtsov} A.~V., 2013, \apj, 766, 25

\bibitem[{{Dolag} {et~al}\mbox{.}(2009){Dolag}, {Borgani}, {Murante}, \&
  {Springel}}]{Dolag2009}
{Dolag} K., {Borgani} S., {Murante} G., {Springel} V., 2009, \mnras, 399, 497

\bibitem[{{Donnari} {et~al}\mbox{.}(2021){Donnari}, {Pillepich}, {Joshi},
  {Nelson}, {Genel}, {Marinacci}, {Rodriguez-Gomez}, {Pakmor}, {Torrey},
  {Vogelsberger}, \& {Hernquist}}]{Donnari2021}
{Donnari} M. {et~al.}, 2021, \mnras, 500, 4004

\bibitem[{{Donnari} {et~al}\mbox{.}(2019){Donnari}, {Pillepich}, {Nelson},
  {Vogelsberger}, {Genel}, {Weinberger}, {Marinacci}, {Springel}, \&
  {Hernquist}}]{Donnari2019}
{Donnari} M. {et~al.}, 2019, \mnras, 485, 4817

\bibitem[{{Dressler}(1980)}]{Dressler1980}
{Dressler} A., 1980, \apj, 236, 351

\bibitem[{{Fattahi} {et~al}\mbox{.}(2018){Fattahi}, {Navarro}, {Frenk}, {Oman},
  {Sawala}, \& {Schaller}}]{Fattahi2018}
{Fattahi} A., {Navarro} J.~F., {Frenk} C.~S., {Oman} K.~A., {Sawala} T.,
  {Schaller} M., 2018, \mnras, 476, 3816

\bibitem[{{Finn} {et~al}\mbox{.}(2023){Finn}, {Vulcani}, {Rudnick}, {Balogh},
  {Desai}, {Jablonka}, \& {Zaritsky}}]{Finn2023}
{Finn} R.~A., {Vulcani} B., {Rudnick} G., {Balogh} M.~L., {Desai} V.,
  {Jablonka} P., {Zaritsky} D., 2023, \mnras, 521, 4614

\bibitem[{{Furlong} {et~al}\mbox{.}(2015){Furlong}, {Bower}, {Theuns},
  {Schaye}, {Crain}, {Schaller}, {Dalla Vecchia}, {Frenk}, {McCarthy}, {Helly},
  {Jenkins}, \& {Rosas-Guevara}}]{Furlong2015}
{Furlong} M. {et~al.}, 2015, \mnras, 450, 4486

\bibitem[{{Gallazzi} {et~al}\mbox{.}(2005){Gallazzi}, {Charlot}, {Brinchmann},
  {White}, \& {Tremonti}}]{Gallazzi2005}
{Gallazzi} A., {Charlot} S., {Brinchmann} J., {White} S. D.~M., {Tremonti}
  C.~A., 2005, \mnras, 362, 41

\bibitem[{{Genel} {et~al}\mbox{.}(2018){Genel}, {Nelson}, {Pillepich},
  {Springel}, {Pakmor}, {Weinberger}, {Hernquist}, {Naiman}, {Vogelsberger},
  {Marinacci}, \& {Torrey}}]{Genal2018}
{Genel} S. {et~al.}, 2018, \mnras, 474, 3976

\bibitem[{{Giocoli}, {Tormen} \& {van den Bosch}(2008){Giocoli}, {Tormen}, \&
  {van den Bosch}}]{Giocoli2008}
{Giocoli} C., {Tormen} G., {van den Bosch} F.~C., 2008, \mnras, 386, 2135

\bibitem[{{Gregory} \& {Thompson}(1978)}]{Gregory1978}
{Gregory} S.~A., {Thompson} L.~A., 1978, \apj, 222, 784

\bibitem[{{Gunn} \& {Gott}(1972)}]{Gunn1972}
{Gunn} J.~E., {Gott}, J.~Richard I., 1972, \apj, 176, 1

\bibitem[{{Harris} {et~al}\mbox{.}(2020){Harris}, {Millman}, {van der Walt},
  {Gommers}, {Virtanen}, {Cournapeau}, {Wieser}, {Taylor}, {Berg}, {Smith},
  {Kern}, {Picus}, {Hoyer}, {van Kerkwijk}, {Brett}, {Haldane}, {del R{\'\i}o},
  {Wiebe}, {Peterson}, {G{\'e}rard-Marchant}, {Sheppard}, {Reddy}, {Weckesser},
  {Abbasi}, {Gohlke}, \& {Oliphant}}]{Harris2020}
{Harris} C.~R. {et~al.}, 2020, \nat, 585, 357

\bibitem[{{Hartzenberg} {et~al}\mbox{.}(2023){Hartzenberg}, {Cowley},
  {Hopkins}, \& {Allen}}]{Hartzenberg2023}
{Hartzenberg} G.~R., {Cowley} M.~J., {Hopkins} A.~M., {Allen} R.~J., 2023,
  \pasa, 40, e043

\bibitem[{{Hough} {et~al}\mbox{.}(2023){Hough}, {Cora}, {Haggar},
  {Vega-Martinez}, {Kuchner}, {Pearce}, {Gray}, {Knebe}, \&
  {Yepes}}]{Hough2023}
{Hough} T. {et~al.}, 2023, \mnras, 518, 2398

\bibitem[{{Hunter}(2007)}]{Hunter2007}
{Hunter} J.~D., 2007, Computing in Science and Engineering, 9, 90

\bibitem[{{Huss}, {Jain} \& {Steinmetz}(1999){Huss}, {Jain}, \&
  {Steinmetz}}]{Huss1999}
{Huss} A., {Jain} B., {Steinmetz} M., 1999, \apj, 517, 64

\bibitem[{{Joshi}, {Wadsley} \& {Parker}(2017){Joshi}, {Wadsley}, \&
  {Parker}}]{Joshi2017}
{Joshi} G.~D., {Wadsley} J., {Parker} L.~C., 2017, \mnras, 468, 4625

\bibitem[{{Knebe} {et~al}\mbox{.}(2011){Knebe}, {Libeskind}, {Knollmann},
  {Martinez-Vaquero}, {Yepes}, {Gottl{\"o}ber}, \& {Hoffman}}]{Knebe2011}
{Knebe} A., {Libeskind} N.~I., {Knollmann} S.~R., {Martinez-Vaquero} L.~A.,
  {Yepes} G., {Gottl{\"o}ber} S., {Hoffman} Y., 2011, \mnras, 412, 529

\bibitem[{{Lange} {et~al}\mbox{.}(2015){Lange}, {Driver}, {Robotham}, {Kelvin},
  {Graham}, {Alpaslan}, {Andrews}, {Baldry}, {Bamford}, {Bland-Hawthorn},
  {Brough}, {Cluver}, {Conselice}, {Davies}, {Haeussler}, {Konstantopoulos},
  {Loveday}, {Moffett}, {Norberg}, {Phillipps}, {Taylor},
  {L{\'o}pez-S{\'a}nchez}, \& {Wilkins}}]{Lange2015}
{Lange} R. {et~al.}, 2015, \mnras, 447, 2603

\bibitem[{{Lopes}, {Ribeiro} \& {Brambila}(2024){Lopes}, {Ribeiro}, \&
  {Brambila}}]{Lopes2024}
{Lopes} P. A.~A., {Ribeiro} A. L.~B., {Brambila} D., 2024, \mnras, 527, L19

\bibitem[{{Ludlow} {et~al}\mbox{.}(2023){Ludlow}, {Fall}, {Wilkinson},
  {Schaye}, \& {Obreschkow}}]{Ludlow2023}
{Ludlow} A.~D., {Fall} S.~M., {Wilkinson} M.~J., {Schaye} J., {Obreschkow} D.,
  2023, \mnras, 525, 5614

\bibitem[{{Ludlow} {et~al}\mbox{.}(2019){Ludlow}, {Schaye}, {Schaller}, \&
  {Richings}}]{Ludlow2019}
{Ludlow} A.~D., {Schaye} J., {Schaller} M., {Richings} J., 2019, \mnras, 488,
  L123

\bibitem[{{Marinacci} {et~al}\mbox{.}(2018){Marinacci}, {Vogelsberger},
  {Pakmor}, {Torrey}, {Springel}, {Hernquist}, {Nelson}, {Weinberger},
  {Pillepich}, {Naiman}, \& {Genel}}]{Marinacci2018}
{Marinacci} F. {et~al.}, 2018, \mnras, 480, 5113

\bibitem[{{More}, {Diemer} \& {Kravtsov}(2015){More}, {Diemer}, \&
  {Kravtsov}}]{More2015}
{More} S., {Diemer} B., {Kravtsov} A.~V., 2015, \apj, 16

\bibitem[{{Mosleh} {et~al}\mbox{.}(2020){Mosleh}, {Hosseinnejad},
  {Hosseini-ShahiSavandi}, \& {Tacchella}}]{Mosleh2020}
{Mosleh} M., {Hosseinnejad} S., {Hosseini-ShahiSavandi} S.~Z., {Tacchella} S.,
  2020, \apj, 905, 170

\bibitem[{{Naiman} {et~al}\mbox{.}(2018){Naiman}, {Pillepich}, {Springel},
  {Ramirez-Ruiz}, {Torrey}, {Vogelsberger}, {Pakmor}, {Nelson}, {Marinacci},
  {Hernquist}, {Weinberger}, \& {Genel}}]{Naiman2018}
{Naiman} J.~P. {et~al.}, 2018, \mnras, 477, 1206

\bibitem[{{Nanni} {et~al}\mbox{.}(2023){Nanni}, {Neumann}, {Thomas},
  {Maraston}, {Trayford}, {Lovell}, {Law}, {Yan}, \& {Chen}}]{Nanni2023}
{Nanni} L. {et~al.}, 2023, arXiv e-prints, arXiv:2309.14257

\bibitem[{{Nelson} {et~al}\mbox{.}(2018){Nelson}, {Pillepich}, {Springel},
  {Weinberger}, {Hernquist}, {Pakmor}, {Genel}, {Torrey}, {Vogelsberger},
  {Kauffmann}, {Marinacci}, \& {Naiman}}]{Nelson2018}
{Nelson} D. {et~al.}, 2018, \mnras, 475, 624

\bibitem[{{Nelson} {et~al}\mbox{.}(2019){Nelson}, {Springel}, {Pillepich},
  {Rodriguez-Gomez}, {Torrey}, {Genel}, {Vogelsberger}, {Pakmor}, {Marinacci},
  {Weinberger}, {Kelley}, {Lovell}, {Diemer}, \& {Hernquist}}]{Nelson2019}
{Nelson} D. {et~al.}, 2019, Computational Astrophysics and Cosmology, 6, 2

\bibitem[{{O'Neil} {et~al}\mbox{.}(2021){O'Neil}, {Barnes}, {Vogelsberger}, \&
  {Diemer}}]{O'Neil2021}
{O'Neil} S., {Barnes} D.~J., {Vogelsberger} M., {Diemer} B., 2021, \mnras, 504,
  4649

\bibitem[{{O'Neil} {et~al}\mbox{.}(2022){O'Neil}, {Borrow}, {Vogelsberger}, \&
  {Diemer}}]{O'Neil2022}
{O'Neil} S., {Borrow} J., {Vogelsberger} M., {Diemer} B., 2022, \mnras, 513,
  835

\bibitem[{{Oyarzun} {et~al}\mbox{.}(2023){Oyarzun}, {Bundy}, {Westfall},
  {Lacerna}, {Yan}, {Brownstein}, {Drory}, \& {Lane}}]{Oyarzun2023}
{Oyarzun} G.~A., {Bundy} K., {Westfall} K.~B., {Lacerna} I., {Yan} R.,
  {Brownstein} J.~R., {Drory} N., {Lane} R.~R., 2023, arXiv e-prints,
  arXiv:2302.12268

\bibitem[{{Paccagnella} {et~al}\mbox{.}(2016){Paccagnella}, {Vulcani},
  {Poggianti}, {Moretti}, {Fritz}, {Gullieuszik}, {Couch}, {Bettoni}, {Cava},
  {D'Onofrio}, \& {Fasano}}]{Paccagnella2016}
{Paccagnella} A. {et~al.}, 2016, \apjl, 816, L25

\bibitem[{{Pakmor} {et~al}\mbox{.}(2016){Pakmor}, {Volker}, {Bauer}, {Mocz},
  {Munoz}, {Ohlmann}, {Schaal}, \& {Zhu}}]{Pakmor2016}
{Pakmor} R., {Volker} S., {Bauer} A., {Mocz} P., {Munoz} D.~J., {Ohlmann}
  S.~T., {Schaal} K., {Zhu} C., 2016, \mnras, 455, 1134

\bibitem[{{P{\'e}rez-Mill{\'a}n} {et~al}\mbox{.}(2023){P{\'e}rez-Mill{\'a}n},
  {Fritz}, {Gonz{\'a}lez-L{\'o}pezlira}, {Moretti}, {Cervantes Sodi},
  {Vulcani}, {Gullieuszik}, {Bruzual}, {Charlot}, \& {Bettoni}}]{Perez2023}
{P{\'e}rez-Mill{\'a}n} D. {et~al.}, 2023, \mnras, 521, 1292

\bibitem[{{Pillepich} {et~al}\mbox{.}(2018{\natexlab{a}}){Pillepich}, {Nelson},
  {Hernquist}, {Springel}, {Pakmor}, {Torrey}, {Weinberger}, {Genel}, {Naiman},
  {Marinacci}, \& {Vogelsberger}}]{Pillepich2018b}
{Pillepich} A. {et~al.}, 2018{\natexlab{a}}, \mnras, 475, 648

\bibitem[{{Pillepich} {et~al}\mbox{.}(2019){Pillepich}, {Nelson}, {Springel},
  {Pakmor}, {Torrey}, {Weinberger}, {Vogelsberger}, {Marinacci}, {Genel}, {van
  der Wel}, \& {Hernquist}}]{Pillepich2019}
{Pillepich} A. {et~al.}, 2019, \mnras, 490, 3196

\bibitem[{{Pillepich} {et~al}\mbox{.}(2018{\natexlab{b}}){Pillepich},
  {Springel}, {Nelson}, {Genel}, {Naiman}, {Pakmor}, {Hernquist}, {Torrey},
  {Vogelsberger}, {Weinberger}, \& {Marinacci}}]{Pillepich2018a}
{Pillepich} A. {et~al.}, 2018{\natexlab{b}}, \mnras, 473, 4077

\bibitem[{{Planck Collaboration} {et~al}\mbox{.}(2016){Planck Collaboration},
  {Ade}, {Aghanim}, {Arnaud}, {Ashdown}, {Aumunt}, {Baccigalupi}, {Banday}, \&
  et~al.}]{PlanckCollaborationXIII2016}
{Planck Collaboration} {et~al.}, 2016, \aap, 594, 63

\bibitem[{{Power} {et~al}\mbox{.}(2003){Power}, {Navarro}, {Jenkins}, {Frenk},
  {White}, {Springel}, {Stadel}, \& {Quinn}}]{Power2003}
{Power} C., {Navarro} J.~F., {Jenkins} A., {Frenk} C.~S., {White} S.~D.~M.,
  {Springel} V., {Stadel} J., {Quinn} T., 2003, \mnras, 338, 14

\bibitem[{{Rodriguez} {et~al}\mbox{.}(2021){Rodriguez}, {Montero-Dorta},
  {Angulo}, {Artale}, \& {Merch{\'a}n}}]{Rodriguez2021}
{Rodriguez} F., {Montero-Dorta} A.~D., {Angulo} R.~E., {Artale} M.~C.,
  {Merch{\'a}n} M., 2021, \mnras, 505, 3192

\bibitem[{{Ruiz} {et~al}\mbox{.}(2023){Ruiz}, {Mart{\'\i}nez}, {Coenda},
  {Muriel}, {Cora}, {de los Rios}, \& {Vega-Mart{\'\i}nez}}]{Ruiz2023}
{Ruiz} A.~N., {Mart{\'\i}nez} H.~J., {Coenda} V., {Muriel} H., {Cora} S.~A.,
  {de los Rios} M., {Vega-Mart{\'\i}nez} C.~A., 2023, \mnras, 525, 3048

\bibitem[{{Sarkar}, {Pandey} \& {Sarkar}(2023){Sarkar}, {Pandey}, \&
  {Sarkar}}]{Sarkar2023}
{Sarkar} P., {Pandey} B., {Sarkar} S., 2023, \mnras, 519, 3227

\bibitem[{{Schaye} {et~al}\mbox{.}(2015){Schaye}, {Crain}, {Bower}, {Furlong},
  {Schaller}, {Theuns}, {Dalla Vecchia}, {Frenk}, {McCarthy}, {Helly},
  {Jenkins}, {Rosas-Guevara}, {White}, {Baes}, {Booth}, {Camps}, {Navarro},
  {Qu}, {Rahmati}, {Sawala}, {Thomas}, \& {Trayford}}]{Schaye2015}
{Schaye} J. {et~al.}, 2015, \mnras, 446, 521

\bibitem[{{Shin} {et~al}\mbox{.}(2019){Shin}, {Adhikari}, {Baxter}, {Chang},
  {Battaglia}, {Bleem}, {Bocquet}, {DeRose}, {Gruen}, {Hilton}, {Kravtsov},
  {McClintock}, {Rozo}, {Rykoff}, {Varga}, {Wechsler}, {Wu}, {Zhang}, {Aiola},
  {Allam}, {Bechtol}, {Benson}, {Bertin}, {Bond}, {Brodwin}, {Brooks},
  {Buckley-Geer}, {Burke}, {Carlstrom}, {Carrasco Kind}, {Carretero},
  {Castander}, {Choi}, {Cunha}, {Crawford}, {da Costa}, {De Vicente}, {Desai},
  {Devlin}, {Dietrich}, {Doel}, {Dunkley}, {Eifler}, {Evrard}, {Flaugher}, P.,
  {Gallardo}, {Garc\'{i}a-Bellido}, {Gaztanaga}, {Gerdes}, {Gralla}, {Gruendl},
  {Gschwend}, {Gitierrez}, {Hartley}, {Hill}, {Hollowood}, {Hoyle},
  {Huffenberger}, {Hughes}, {James}, {Jeltema}, {Kim}, {Krause}, {Kuehn},
  {Lahav}, {Lima}, {Madhavacheril}, {Maia}, {Marshall}, {Maurin}, {McMahon},
  {Menanteau}, {Miller}, {Miquel}, {Mohr}, {Naess}, {Nati}, {Newburgh},
  {Niemack}, {Ogando}, {Partridge}, {Patil}, {Plazas}, {Rapetti}, {Reichardt},
  {Romer}, {Sanchez}, {Scarpine}, {Schindler}, {Serrano}, {Smith}, {Smith},
  {Soares-Santos}, {Sobreira}, {Staggs}, {Stark}, {Stein}, E., {Swanson},
  {Tarle}, {Thomas}, {van Engelen}, {Wollack}, \& {Xu}}]{Shin2019}
{Shin} T. {et~al.}, 2019, \mnras, 487, 2900

\bibitem[{{Speagle} {et~al}\mbox{.}(2014){Speagle}, {Steinhardt}, {Capak}, \&
  {Silverman}}]{Speagle2014}
{Speagle} J.~S., {Steinhardt} C.~L., {Capak} P.~L., {Silverman} J.~D., 2014,
  \apjs, 214, 15

\bibitem[{{Springel}(2005)}]{Springel2005b}
{Springel} V., 2005, \mnras, 364, 1105

\bibitem[{{Springel}(2010)}]{Springel2010}
{Springel} V., 2010, \mnras, 401, 791

\bibitem[{{Springel} \& {Hernquist}(2003)}]{Springel2003}
{Springel} V., {Hernquist} L., 2003, \mnras, 339, 289

\bibitem[{{Springel} {et~al}\mbox{.}(2018){Springel}, {Pakmor}, {Pillepich},
  {Weinberger}, {Nelson}, {Hernquist}, {Vogelsberger}, {Genel}, {Torrey},
  {Marinacci}, \& {Naiman}}]{Springel2018}
{Springel} V. {et~al.}, 2018, \mnras, 475, 676

\bibitem[{{Springel} {et~al}\mbox{.}(2001){Springel}, {White}, {Tormen}, \&
  {Kauffmann}}]{Springel2001}
{Springel} V., {White} S.~D.~M., {Tormen} G., {Kauffmann} G., 2001, \mnras,
  328, 726

\bibitem[{{Torrey} {et~al}\mbox{.}(2019){Torrey}, {Vogelsberger}, {Marinacci},
  {Pakmor}, {Springel}, {Nelson}, {Naiman}, {Pillepich}, {Genel}, {Weinberger},
  \& {Hernquist}}]{Torrey2019}
{Torrey} P. {et~al.}, 2019, \mnras, 484, 5587

\bibitem[{{van den Bosch} {et~al}\mbox{.}(2016){van den Bosch}, {Fangzhou},
  {Campbell}, \& {Behroozi}}]{vandenBosch2016}
{van den Bosch} F.~C., {Fangzhou} J., {Campbell} D., {Behroozi} P., 2016,
  \mnras, 158

\bibitem[{{van den Bosch} {et~al}\mbox{.}(2008){van den Bosch}, {Pasquali},
  {Yang}, {Mo}, {Weinmann}, {McIntosh}, \& {Aquino}}]{vandenBosch2008}
{van den Bosch} F.~C., {Pasquali} A., {Yang} X., {Mo} H.~J., {Weinmann} S.,
  {McIntosh} D.~H., {Aquino} D., 2008, arXiv e-prints, arXiv:0805.0002

\bibitem[{{van der Burg} {et~al}\mbox{.}(2020){van der Burg}, {Rudnick},
  {Balogh}, {Muzzin}, {Lidman}, {Old}, {Shipley}, {Gilbank}, {McGee},
  {Biviano}, {Cerulo}, {Chan}, {Cooper}, {De Lucia}, {Demarco}, {Forrest},
  {Gwyn}, {Jablonka}, {Kukstas}, {Marchesini}, {Nantais}, {Noble},
  {Pintos-Castro}, {Poggianti}, {Reeves}, {Stefanon}, {Vulcani}, {Webb},
  {Wilson}, {Yee}, \& {Zaritsky}}]{vanderBurg2020}
{van der Burg} R. F.~J. {et~al.}, 2020, \aap, 638, A112

\bibitem[{Van~Rossum \& Drake~Jr(1995)}]{vanRossum1995}
Van~Rossum G., Drake~Jr F.~L., 1995, Python reference manual. Centrum voor
  Wiskunde en Informatica Amsterdam

\bibitem[{{Virtanen} {et~al}\mbox{.}(2020){Virtanen}, {Gommers}, {Oliphant},
  {Haberland}, {Reddy}, {Cournapeau}, {Burovski}, {Peterson}, {Weckesser},
  {Bright}, {van der Walt}, {Brett}, {Wilson}, {Millman}, {Mayorov}, {Nelson},
  {Jones}, {Kern}, {Larson}, {Carey}, {Polat}, {Feng}, {Moore}, {VanderPlas},
  {Laxalde}, {Perktold}, {Cimrman}, {Henriksen}, {Quintero}, {Harris},
  {Archibald}, {Ribeiro}, {Pedregosa}, {van Mulbregt}, \& {SciPy 1. 0
  Contributors}}]{Virtanen2020}
{Virtanen} P. {et~al.}, 2020, Nature Methods, 17, 261

\bibitem[{{Vogelsberger} {et~al}\mbox{.}(2014{\natexlab{a}}){Vogelsberger},
  {Genel}, {Springel}, {Torrey}, {Sijacki}, {Xu}, {Snyder}, {Bird}, {Nelson},
  \& {Hernquist}}]{Vogelsberger2014a}
{Vogelsberger} M. {et~al.}, 2014{\natexlab{a}}, \nat, 509, 177

\bibitem[{{Vogelsberger} {et~al}\mbox{.}(2014{\natexlab{b}}){Vogelsberger},
  {Genel}, {Springel}, {Torrey}, {Sijacki}, {Xu}, {Snyder}, {Nelson}, \&
  {Hernquist}}]{Vogelsberger2014b}
{Vogelsberger} M. {et~al.}, 2014{\natexlab{b}}, \mnras, 444, 1518

\bibitem[{{Vogelsberger} {et~al}\mbox{.}(2018){Vogelsberger}, {Marinacci},
  {Torrey}, {Genel}, {Springel}, {Weinberger}, {Pakmor}, {Hernquist}, {Naiman},
  {Pillepich}, \& {Nelson}}]{Vogelsberger2018}
{Vogelsberger} M. {et~al.}, 2018, \mnras, 474, 2073

\bibitem[{{Wang} {et~al}\mbox{.}(2018){Wang}, {Wang}, {Mo}, {Lim}, {van den
  Bosch}, {Kong}, {Wang}, {Yang}, \& {Chen}}]{Wang2018}
{Wang} E. {et~al.}, 2018, \apj, 860, 102

\bibitem[{{Weinberger} {et~al}\mbox{.}(2017){Weinberger}, {Springel},
  {Hernquist}, {Pillepich}, {Marinacci}, {Pakmor}, {Nelson}, {Genel},
  {Vogelsberger}, {Naiman}, \& {Torrey}}]{Weinberger2017}
{Weinberger} R. {et~al.}, 2017, \mnras, 465, 3291

\bibitem[{{Weinberger}, {Springel} \& {Pakmor}(2020){Weinberger}, {Springel},
  \& {Pakmor}}]{Weinberger2020}
{Weinberger} R., {Springel} V., {Pakmor} R., 2020, The Astrophysical Journal
  Supplement Series, 248, 39

\bibitem[{{Whitaker} {et~al}\mbox{.}(2014){Whitaker}, {Franx}, {Leja}, {van
  Dokkum}, {Henry}, {Skelton}, {Fumagalli}, {Momcheva}, {Brammer}, {Labb{\'e}},
  {Nelson}, \& {Rigby}}]{Whitaker2014}
{Whitaker} K.~E. {et~al.}, 2014, \apj, 795, 104

\bibitem[{{Xhakaj} {et~al}\mbox{.}(2020){Xhakaj}, {Diemer}, {Leauthaud},
  {Wasserman}, {Huang}, {Luo}, {Adhikari}, \& {Singh}}]{Xhakaj2020}
{Xhakaj} E., {Diemer} B., {Leauthaud} A., {Wasserman} A., {Huang} S., {Luo} Y.,
  {Adhikari} S., {Singh} S., 2020, \mnras, 499, 3534

\end{thebibliography}

\appendix

\section{Investigating spurious mass assignment in the galaxy identification algorithm}
\label{ap:radius}

\begin{figure*}
  \centering
  \includegraphics[]{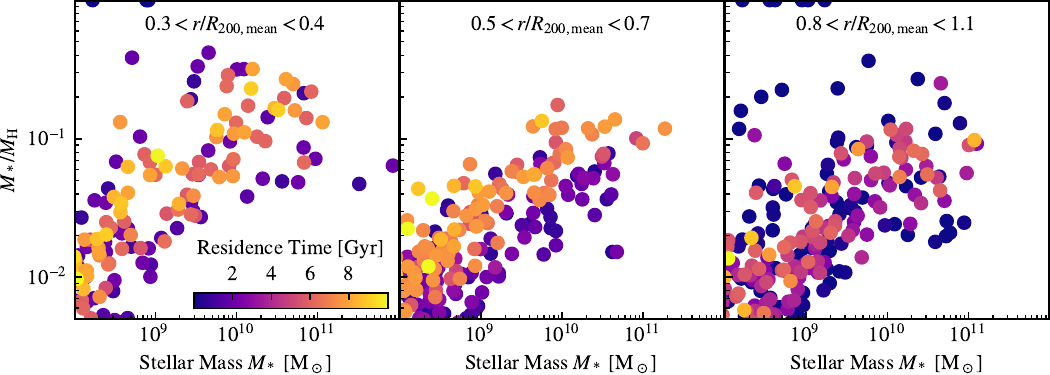}
  \caption{The stellar to halo mass ratio as a function of stellar mass for the most massive halo in the volume. Each panel shows a selection of galaxies from a fixed radius range, as a function of $R_{\rm 200, mean}$, with the points coloured by the galaxies' residence time in the cluster.  The primary contribution to the change in halo mass, shown in Figures \ref{fig:smhmrresidence}-\ref{fig:mass_smhm}, is the residence time of galaxies.  There is not a major difference in the distribution at different radii (i.e. in each panel), although there is a slight increase in $M_\star/M_{\rm H}$ for galaxies at small radii.}
  \label{fig:ratio_residence_time}
\end{figure*}

In Figure \ref{fig:ratio_residence_time}, we show the stellar to halo mass ratio as a function of the stellar mass of galaxies, for the most massive cluster in our sample ($M_{\rm 200, mean} = 2.1 \times 10^{15}$ M$_\odot$).  We split each panel by galaxy radius within the cluster at redshift $z=0$, and colour each point by the residence time of the galaxy.

We visualise the data in this way to explicitly show that the major contributor to the offset (vertically, i.e. the reduction of halo mass) is indeed the residence time of the galaxies. Galaxies that have resided in the cluster for a longer time have demonstrably higher stellar to halo mass ratios.

The galaxies that reside closer in to the centre of the cluster (with $r / R_{\rm 200, mean} < 0.4$) also see a systematic vertical offset here. It is unclear whether this is driven by the physical (more dense) environment, or a systematic under-estimation of the halo mass by the halo and galaxy finder due to denser surroundings. 

\begin{figure}
    \centering
    \includegraphics{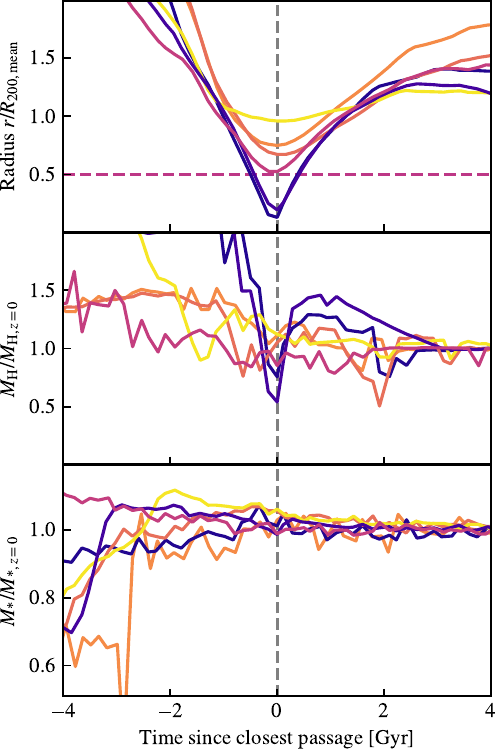}
    \caption{Six randomly selected galaxies from the most massive cluster in the TNG300 volume, with tracks showing their radius relative to the center (note that lines are coloured based upon the distance of closest passage), the halo mass of the galaxies as computed by {\tt SubFind} relative to its value at $z=0$, and the galaxy stellar mass (the stellar mass within twice the half-mass radius), again relative to its value at $z=0$. A horizontal dashed line in the top panel shows the limiting case of $r = 0.5 R_{\rm 200, mean}$ where there appears to be significant numerical errors in the computation of the halo mass, and the grey vertical dashed line shows the time of closest passage for each galaxy.}
    \label{fig:infall_supression}
\end{figure}

In Figure \ref{fig:infall_supression} we attempt to further disentangle the impact of potential halo finder errors on the computation of halo masses. We show, from the top panel down, the orbital radius of each galaxy as a function of time (relative to their time of closest passage to the BCG), their halo mass, and their galaxy stellar mass.

First, it is clear from this figure that as the galaxies enter the cluster (about 1-2 Gyr before their closest passage), they see both a systematic loss of bound halo mass (which remains even after they leave the cluster environment again at a later time), and are quenched (i.e. their stellar mass stops increasing, and remains flat). We can also identify that haloes with closer passages lose a significantly higher fraction of their bound mass.

Secondly, we see a potentially spurious numerical feature in the galaxies that have the closest passage with the central galaxy (the darkest lines in the figure). These galaxies see a strong suppression in the halo mass once they enter a sphere with radius $r = 0.5 R_{\rm 200, mean}$. This is likely due to their haloes having some of their bound component temporarily identified as bound to the central galaxy. Once the haloes leave this sphere again, this spurious misidentification ends. Thankfully, we see no such spurious impact on the stellar properties of galaxies, due to this being in a much denser portion of the subhalo and hence much less susceptible to these effects.

\begin{figure}
    \centering
    \includegraphics{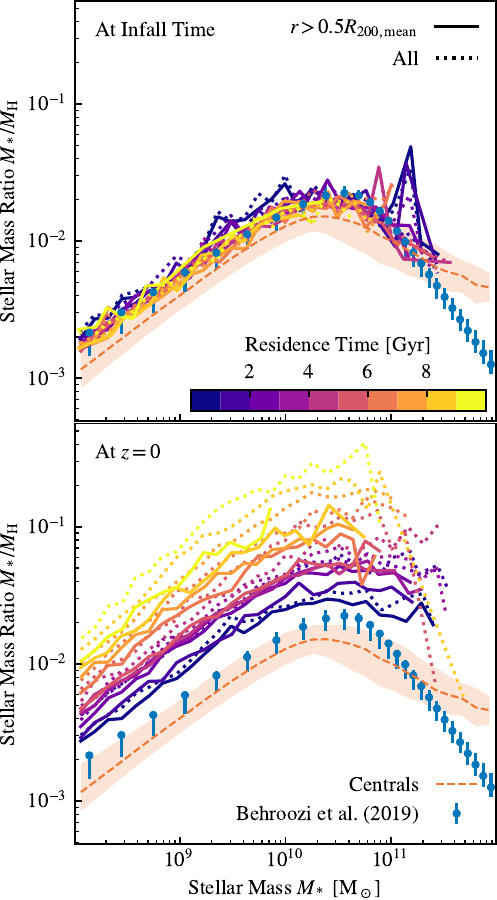}
    \caption{Same as Figure \ref{fig:mass_smhm}, but now only including galaxies that live at radii $r > 0.5 R_{\rm 200, mean}$ as motivated by Figure \ref{fig:infall_supression}.  The galaxy population is split by residence time and we plot the stellar mass ratio as a function of stellar mass, with each residence time corresponding to a different colour. Dotted lines show all galaxies (i.e. the exact same lines as Figure \ref{fig:mass_smhm}), with solid lines including the new cut of $r > 0.5 R_{\rm 200, mean}$.  While there is a slight decrease in $M_\star/M_{H}$ when excluding the galaxies at smaller radii, the systematic increase with residence time remains. }
    \label{fig:smhmr_0p5}
\end{figure}

In Figure \ref{fig:smhmr_0p5}, we show the stellar to halo mass ratio again, as in Figure \ref{fig:mass_smhm}, but now excluding all galaxies that lie at $z=0$ within $r < 0.5 R_{\rm 200, mean}$ from the brightest cluster galaxy. This should remove any galaxies that have spuriously low halo masses (and hence systematically high $M_\star / M_{\rm H}$ ratios) due to their current dense environment and proximity to the BCG. The trend of galaxies that have resided within the cluster for longer having higher stellar to halo mass ratios remains, and hence gives us further confidence that our results here are primarily driven by the loss of halo mass from the extremities of galaxies.

\label{lastpage}

\end{document}